\documentclass[twocolumn]{aastex63}

\usepackage{graphicx}
\usepackage[caption=false]{subfig}

\usepackage{xspace}

\newcommand\asec{$^{\prime\prime}$\xspace}

\begin{document}

\title{Identification of hot gas around low-mass protostars}

\correspondingauthor{Merel L.R. van 't Hoff}
\email{mervth@umich.edu}

\author[0000-0002-2555-9869]{Merel L.R. van 't Hoff}
\affil{Department of Astronomy, University of Michigan, 1085 S. University Ave., Ann Arbor, MI 48109-1107, USA}

\author[0000-0003-4179-6394]{Edwin A. Bergin}
\affil{Department of Astronomy, University of Michigan, 1085 S. University Ave., Ann Arbor, MI 48109-1107, USA}

\author{Penelope Riley}
\affil{Department of Physics and Astrophysics, DePaul University, Chicago, IL 60614-3504, USA}
\affil{Department of Astronomy, University of Illinois, 1002 West Green St., Urbana, IL 61801, USA}

\author{Sanil Mittal}
\affil{Department of Astronomy, University of Michigan, 1085 S. University Ave., Ann Arbor, MI 48109-1107, USA}

\author[0000-0001-9133-8047]{Jes K. J{\o}rgensen}
\affil{Niels Bohr Institute, University of Copenhagen, {\O}ster Voldgade 5--7, DK-1350, Copenhagen K, Denmark}

\author[0000-0002-6195-0152]{John J. Tobin}
\affil{National Radio Astronomy Observatory, 520 Edgemont Rd., Charlottesville, VA 22903, USA}


\begin{abstract} 
The low carbon content of Earth and primitive meteorites compared to the Sun and interstellar grains suggests that carbon-rich grains were destroyed in the inner few astronomical units of the young solar system. A promising mechanism to selectively destroy carbonaceous grains is thermal sublimation within the soot line at $\gtrsim$ 300 K. To address whether such hot conditions are common amongst low-mass protostars, we observe CH$_3$CN transitions at 1, 2 and 3 mm with the NOrthern Extended Millimeter Array (NOEMA) toward seven low-mass and one intermediate-mass protostar ($L_{\rm{bol}} \sim2-300 L_\odot$), as CH$_3$CN is an excellent temperature tracer. We find $>$ 300~K gas toward all sources, indicating that hot gas may be prevalent. Moreover, the excitation temperature for CH$_3$OH obtained with the same observations is always lower ($\sim$135--250 K), suggesting that CH$_3$CN and CH$_3$OH have a different spatial distribution. A comparison of the column densities at 1 and 3 mm shows a stronger increase at 3 mm for CH$_3$CN than for CH$_3$OH. Since the dust opacity is lower at longer wavelengths, this indicates that CH$_3$CN is enhanced in the hot gas compared to CH$_3$OH. If this CH$_3$CN enhancement is the result of carbon-grain sublimation, these results suggests that Earth's initial formation conditions may not be rare. 
\vspace{1cm}
\end{abstract}


\section{Introduction}

One aspect of determining the occurrence rate of Earth-like planets is understanding the conditions in the early solar system and the prevalence of these conditions amongst low-mass protostars. The composition of Earth can provide information about the composition of the material from which it formed, which is related to the physical conditions in the young disk. For example, since Earth is water poor, it must have formed inside the water snowline. Carbon is also informative as Earth is at least two orders of magnitude depleted in carbon compared to the Sun, interstellar grains and the two well-characterized comets 1P/Halley and 67P/Churyumov-Gerasimenko \citep{Geiss1987,Bergin2015,Rubin2019,Li2021}. A similar amount of depletion is seen in CI chondrites, which represent the most primitive material in the solar system and otherwise reflect solar abundances \citep{Wasson1988}. This means that carbon had to be predominantly present in the gas phase instead of in the refractory phase in the inner young solar system, and as such, could not be efficiently accreted onto forming rocky bodies. 

Oxidation processes can selectively destroy carbonaceous grains while leaving silicate grains intact, but detailed models suggest that these processes by themselves are ineffective \citep{Anderson2017,Klarmann2018,Binkert2023}. An alternative (additional) mechanism is the thermal sublimation of carbon-rich grains inside the ``soot line'' \citep{Kress2010,Li2021}. The exact temperature at which this sublimation occurs depends on the form of the refractory carbon, with the dominant carbon carrier ($\sim$60\%), organic material, sublimating at temperatures between $\sim$350--450 K \citep{Nakano2003,Gail2017}. Constraints can also be placed based on the composition of meteorites \citep{Li2021}: meteoritic materials carry more carbon than water \citep{Hirschmann2009}, but less carbon than sulfur, with C/S ratios well below solar \citep{Hirschmann2016}. This suggests that the sublimation temperature of the main carbon carrier is between that of water ($\sim$150--200 K; \citealt{Fraser2001}) and the main sulfur carrier, troilite (680 K; \citealt{Pollack1994}). Both results then suggest a sublimation temperature of roughly 400 K, compared to $\sim$1200 K for silicates \citep{Pollack1994}. 

Multiple lines of evidence suggest that carbon-grain sublimation must be active during the protostellar phase. First, grain growth already starts while the disk is still forming within an infalling envelope \citep[e.g.,][]{Segura-Cox2020,Tychoniec2020}, and smaller bodies are more easy to sublimate than larger bodies. Second, high temperatures are more easily reached during the protostellar phase when stellar accretion rates are higher and accretion bursts are more frequent. \citet{Li2021} demonstrated that temperatures were only high enough for carbon-grain sublimation at 1 au within the first couple hundred-thousand years. Since sublimation has a well characterized exponential relation between pressure and temperature, and the density in the inner region of protostellar envelopes is lower compared to the inner disk, carbon-grain sublimation is expected to occur at temperatures of $\gtrsim$300 K around protostars \citep{vantHoff2020}.  

\citet{vantHoff2020} argue that the destruction of carbon-rich grains will lead to the gas-phase formation of hydrocarbons and nitriles (molecules with a C$\equiv$N bond, such as HCN and CH$_3$CN) inside the soot line  (see also the chemical models by \citealt{Wei2019}) as, in addition to carbon, a significant fraction of the total nitrogen abundance is present in carbonaceous grains. Carbon-grain sublimation is thus expected to enhance the abundance of nitrogen-bearing complex organic molecules (N-COMs) inside the soot line while oxygen-bearing COMs (O-COMs) are expected to be present in the gas phase more uniformly out to the water snowline ($\sim$100--150 K; \citealt{Penteado2017,Ferrero2020}, and references therein). This suggests that we may observe smaller emitting areas and higher excitation temperatures for N-COMs compared to O-COMs. Even in the absence of carbon-grain sublimation, some astrochemical models predict that certain reactions, such as hydrogen abstraction from ammonia, become active at $\gtrsim$ 300 K, altering the chemical composition of hot gas compared to warm gas just inside the water snowline \citep{Garrod2013,Garrod2017,Garrod2022}. 

\begin{deluxetable}{llcccc}
\tablecaption{Overview of sources. \label{tab:OverviewSources}}
\tablewidth{0pt}
\addtolength{\tabcolsep}{-3pt} 
\tabletypesize{\scriptsize}
\tablehead{
\colhead{Source} \vspace{-0.3cm} & \colhead{R.A.$^a$} & \colhead{Dec.$^a$} & \colhead{Class} & \colhead{$L_{\rm{bol}}$} & \colhead{Ref.$^b$}  \\ 
\colhead{} \vspace{-0.5cm}& \colhead{(J2000)} & \colhead{(J2000)} & \colhead{} & \colhead{($L_{\odot}$)} & \colhead{} \\ 
} 
\startdata 
HBC494      & 05:40:27.45  & -07:27:29.67  & I       & $\sim$300 & 1 \\
HOPS-370    & 05:35:27.63  & -05:09:34.42  & 0/I     & 314       & 2 \\
Per-emb-17  & 03:27:39.11  & +30:13:02.93  & 0       & 7.1$^c$   & 3 \\
Per-emb-20  & 03:27:43.28  & +30:12:28.79  & 0/I     & 2.4       & 3 \\
Ser-emb-1   & 18:29:09.09  & +00:31:30.86  & 0       & 4.5       & 4 \\
Ser-emb-8   & 18:29:48.09  & +01:16:43.26  & 0       & 5.9$^d$   & 4 \\
Ser-emb-11E & 18:29:06.68  & +00:30:33:80  & 0/I$^c$ & 5.3$^c$   & 4 \\
Ser-emb-11W & 18:29:06.63  & +00:30:33:90  & 0/I$^c$ & 5.3$^c$   & 4 \\
Ser-emb-17  & 18:29:06.21  & +00:30:42.99  & I       & 4.2       & 4 \\
Ser-emb ALMA 1 & 18:29:05.55 & +00:30:35.14 & ?      & ?         & 5 \\
\enddata
\vspace{-0.2cm}
\tablenotetext{a}{Position of the continuum peak in the 252 GHz datasets.}
\vspace{-0.3cm} 
\tablenotetext{b}{References. (1) \citet{Strom1993} (2) \citet{Furlan2016} (3) \citet{Tobin2016} scaled to a distance of 300 pc for Perseus \citep{Ortiz-Leon2018a} (4) \citet{Enoch2009} scaled to a distance of 436 for Serpens \citep{Ortiz-Leon2018b} (5) \citet{Martin-Domenech2021}.}
\vspace{-0.3cm}
\tablenotetext{c}{For both sources in the binary combined. Per-emb-17 is an unresolved binary in our observations, while we resolve the East and West component of Ser-emb-11.}
\vspace{-0.3cm}
\tablenotetext{d}{Likely an underestimate, because the source is saturated in 70 $\mu$m Spitzer observations \citep{Enoch2009}.}
\end{deluxetable}

A rich hydrocarbon chemistry and high C/O ratio, potentially due to carbon-grain destruction, was recently detected toward a low-mass disk with JWST \citep{Tabone2023}, but observational constraints on the occurrence of carbon-grain sublimation, and of hot gas more generally, around low-mass protostars are scarce. Existing chemical surveys of low-mass protostars often either focus only on O-COMs \citep[e.g.,][]{Bouvier2022} or cover not enough transitions with different upper level energies to derive excitation temperatures \citep[e.g.,][]{Yang2021}. 

Therefore, in order to understand the conditions present during the initial stages of terrestrial-planet formation, we set out to search for hot gas around low-mass protostars and characterize its chemistry. We conducted the NOEMA program CATS-N-DOGS (Chemistry And Temperature: searching with NOEMA for Destruction of carbon GrainS) consisting of 0$\farcs$5--1$\farcs$5 resolution observations of nine low-mass and one intermediate-mass protostar, covering $\sim$16 GHz of bandwidth at 2 MHz resolution at 1, 2 and 3 mm. Here we present the results for the simplest O-COM (methanol; CH$_3$OH) and N-COM (methyl cyanide; CH$_3$CN), showing that hot ($\gtrsim$ 300 K) gas is present toward all sources in our sample and that CH$_3$CN traces hotter gas than CH$_3$OH. 

The paper is structured as follows. The targets and observations are described in Section~\ref{sec:Observations}. The results for CH$_3$CN and CH$_3$OH are presented in Sections~\ref{sec:CH3CN}--\ref{sec:CH313CN} and \ref{sec:CH3OH}, respectively, and the implications of these results are discussed in Section~\ref{sec:Discussion}. Finally, the main conclusions are summarized in Section~\ref{sec:Conclusion}.

\begin{deluxetable*}{@{\extracolsep{4pt}}lrccrccrcc}
\tablecaption{Overview of observations. \label{tab:Observations}}
\tablewidth{0pt}
\addtolength{\tabcolsep}{-2pt} 
\tabletypesize{\scriptsize}
\tablehead{
\colhead{} & \multicolumn{3}{c}{1 mm (CH$_3$CN $J=14-13$)} & 
             \multicolumn{3}{c}{2 mm (CH$_3$CN $J=12-11$)} & 
             \multicolumn{3}{c}{3 mm (CH$_3$CN $J=5-4$)} \\
\cline{2-4} \cline{5-7} \cline{8-10} 
Sources \vspace{-0.3cm} & \colhead{$I_{\rm{peak}}$(cont)$^a$} & \colhead{beam} & \colhead{line rms$^b$} &
                    \colhead{$I_{\rm{peak}}$(cont)$^a$} & \colhead{beam} & \colhead{line rms$^b$} &
                    \colhead{$I_{\rm{peak}}$(cont)$^a$} & \colhead{beam} & \colhead{line rms$^c$} \\
\colhead{} & \colhead{(K)} & \colhead{(arcsec)} & \colhead{(K)} &  
             \colhead{(K)} & \colhead{(arcsec)} & \colhead{(K)} &  
             \colhead{(K)} & \colhead{(arcsec)} & \colhead{(K)}
} 
\startdata 
HBC494      & 7.92 $\pm$ 0.01 & 0.98$\times$0.32 & 0.25  & 
              \multicolumn{1}{c}{\nodata} & \nodata & \nodata  &
              1.288 $\pm$ 0.001 & 2.58$\times$0.88 & 0.14 \\              
HOPS-370    & 1.30 $\pm$ 0.01 & 3.37$\times$0.60 & 0.02 & 
              1.020 $\pm$ 0.010 & 4.18$\times$0.48 & 0.04 &
              \multicolumn{1}{c}{\nodata}  & \nodata  & \nodata \\
Per-emb-17  & 3.54 $\pm$ 0.01 & 0.50$\times$0.34 & 0.26 &
              1.057 $\pm$ 0.003 & 1.30$\times$0.42 & 0.18 &   
              0.370 $\pm$ 0.002 & 1.33$\times$1.09 & 0.19  \\
Per-emb-20  & 0.36 $\pm$ 0.05 & 0.49$\times$0.34 & 0.18 &   
              0.092 $\pm$ 0.003 & 1.39$\times$0.43 & 0.18 &   
              0.055 $\pm$ 0.001 & 1.29$\times$1.04 & 0.19 \\
Ser-emb-1   & 8.74 $\pm$ 0.01 & 0.76$\times$0.32 & 0.19 &   
              2.222 $\pm$ 0.009 & 1.72$\times$0.70 & 0.17 &   
              0.896 $\pm$ 0.002  & 2.32$\times$0.83 & 0.13 \\
Ser-emb-8   & 3.44 $\pm$ 0.02 & 0.75$\times$0.32 & 0.20 &   
              1.084 $\pm$ 0.006 & 1.71$\times$0.71 & 0.17 &   
              0.416 $\pm$ 0.002  & 2.29$\times$0.84 & 0.13 \\
Ser-emb-11E$^{d}$ & 0.30 $\pm$ 0.01 & 0.77$\times$0.33 & 0.19 &   
                    0.110 $\pm$ 0.006 & 1.81$\times$0.70 & 0.16  &   
                    0.200 $\pm$ 0.002 & 2.33$\times$0.83 & 0.13 \\
Ser-emb-11W$^{d}$ & 1.17 $\pm$ 0.01 & 0.77$\times$0.33 & 0.19  &   
                    0.338 $\pm$ 0.006 & 1.81$\times$0.70 & 0.16  &   
                    0.192 $\pm$ 0.002  & 2.33$\times$0.83 & 0.13 \\
Ser-emb-17$^{d}$ & 5.44 $\pm$ 0.01 & 0.77$\times$0.33 & 0.19 &   
                   1.552 $\pm$ 0.006 & 1.81$\times$0.70 & 0.16 &   
                   0.352 $\pm$ 0.002  & 2.33$\times$0.83 & 0.13 \\
Ser-emb ALMA 1$^{d}$ & 0.15 $\pm$ 0.01 & 0.77$\times$0.33 & 0.19  & 
                       0.052 $\pm$ 0.006 & 1.81$\times$0.70 & 0.16  &   
                       0.064 $\pm$ 0.002 & 2.33$\times$0.83 & 0.13 \\
\enddata
\vspace{0.1cm}
\textbf{Notes.} Three dots (\nodata) indicate that a particular setting has not been observed. 
\vspace{-0.2cm}\tablenotetext{a}{Continuum peak flux in the side band containing the CH$_3$CN transition. The central frequencies of these side bands are 89 GHz, 222 GHz and 260 GHz.}
\vspace{-0.2cm}
\tablenotetext{b}{In 2 MHz channels (2.3--2.7 km s$^{-1}$).}
\vspace{-0.3cm}
\tablenotetext{c}{In 2 km s$^{-1}$ channels.}
\vspace{-0.3cm}
\tablenotetext{d}{Present in the same field of view.}
\end{deluxetable*}


\section{Observations and methods}\label{sec:Observations}

\subsection{Source sample}

We target low-mass protostars known to display CH$_3$OH emission \citep{Hsieh2019,Bergner2019}, but are not part of the MPG-IRAM Observatory Programs (MIOP)\footnote{https://iram-institute.org/science-portal/proposals/lp/miop/}: Per-emb-17, Per-emb-20, Ser-emb-1, Ser-emb-8 (also known as S68N) and Ser-emb-17. The field of view of Ser-emb-17 contains the protostellar binary Ser-emb-11 and the protostar Ser-emb ALMA 1 \citep{Martin-Domenech2021}. The sources in Ser-emb-11 are separated by $\sim$2\asec, and only the western source displays emission from complex organics \citep{Martin-Domenech2021}. The same work detected only one weak low-excitation (50 K) CH$_3$OH transition toward Ser-emb ALMA 1 and therefore concluded that the source most likely does not harbor a hot corino. Per-emb-17 is a close binary (0\farcs28; \citealt{Tobin2016}), which we do not resolve. The luminosities for these sources range between 2.4--7.1 $L_\odot$ (Table~\ref{tab:OverviewSources}). In addition we targeted the FU-Ori type protostar HBC494 (also known as Reipurth 50 N IRS 1; \citealt{Reipurth1986}) with a luminosity of $\sim$300 $L_\odot$ \citep{Strom1993}, as this type of luminous sources can display line-rich spectra \citep[e.g.,][]{vantHoff2018,Lee2019}. We also observe the intermediate-mass protostar HOPS-370 (314 $L_\odot$) also known as OMC2-FIR3 \citep{Chini1997} and VLA 11 \citep{Reipurth1999}, for which a protostellar disk and CH$_3$OH have been detected \citep{Tobin2019,Tobin2020}. With its higher luminosity and hence warmer circumstellar material, HOPS-370 may serve as a template for what hot-gas emission could look like. Both HBC494 and HOPS-370 are located in the Orion molecular cloud. Source properties are listed in Table~\ref{tab:OverviewSources}. 

\subsection{Observations}

For all sources, except HBC494 and HOPS-370, we observe CH$_3$CN $J_K=5_K-4_K$ (91 GHz), $J_K=12_K-11_K$ (220 GHz) and $J_K=14_K-13_K$ (257 GHz). $J_K=5_K-4_K$ is not observed toward HOPS-370 and $J_K=12_K-11_K$ is not observed toward HBC494. All three settings cover a suite of CH$_3$OH transitions. The entire dataset is obtained through five different IRAM-NOEMA programs and an observing log is presented in Table~\ref{tab:NOEMA}. All observations are carried out between February 2020 and November 2021 in A configuration, except for the 220 GHz observations toward the Serpens sources, which are obtained with the more compact C configuration. The HOPS-370 257 GHz observations are taken in both A and C configuration. The PolyFiX correlator provides $\sim$16 GHz of bandwidth, divided over two $\sim$8 GHz sidebands, at 2 MHz resolution. In addition, several narrow, 62.5 kHz resolution spectral windows are placed within the sidebands. Our 1 mm setting covers the frequency ranges of 240.9--248.6 GHz and 256.4--264.1 GHz, the 2 mm setting covers the frequency ranges of 202.7--210.8 GHz and 218.2--226.3 GHz, and the 3 mm setting covers the frequency ranges of 85.2--93.3 GHz and 100.7--108.8 GHz. Here we focus on the 2 MHz resolution spectra, except for CH$_3$CN $5_K-4_K$, for which we use a high spectral resolution window between 91.92--92.01 GHz binned to 2 km s$^{-1}$ to obtain a velocity resolution comparable to that of the $12_K-11_K$ and $14_K-13_K$ transitions. 

The data are calibrated using the CLIC package of the GILDAS\footnote{https://www.iram.fr/IRAMFR/GILDAS/} software, using the most recent GILDAS version available at the time of data reduction (July after a dataset was observed, i.e., 2020 or 2021). The imaging and cleaning was done using the MAPPING package of GILDAS. Self-calibration was performed on the continuum, except for Per-emb-20 for which the continuum is too weak and for HOPS-370 for which the spectrum is too line rich to isolate enough line free channels. The self-calibration solutions were transferred to the spectral line cubes, which were then continuum subtracted and cleaned. The resulting beam sizes and rms are listed in Table~\ref{tab:Observations}. Finally, we extracted spectra toward the continuum peak pixel using the CLASS package of GILDAS. 

\begin{figure*}
\centering
\includegraphics[width=0.95\textwidth]{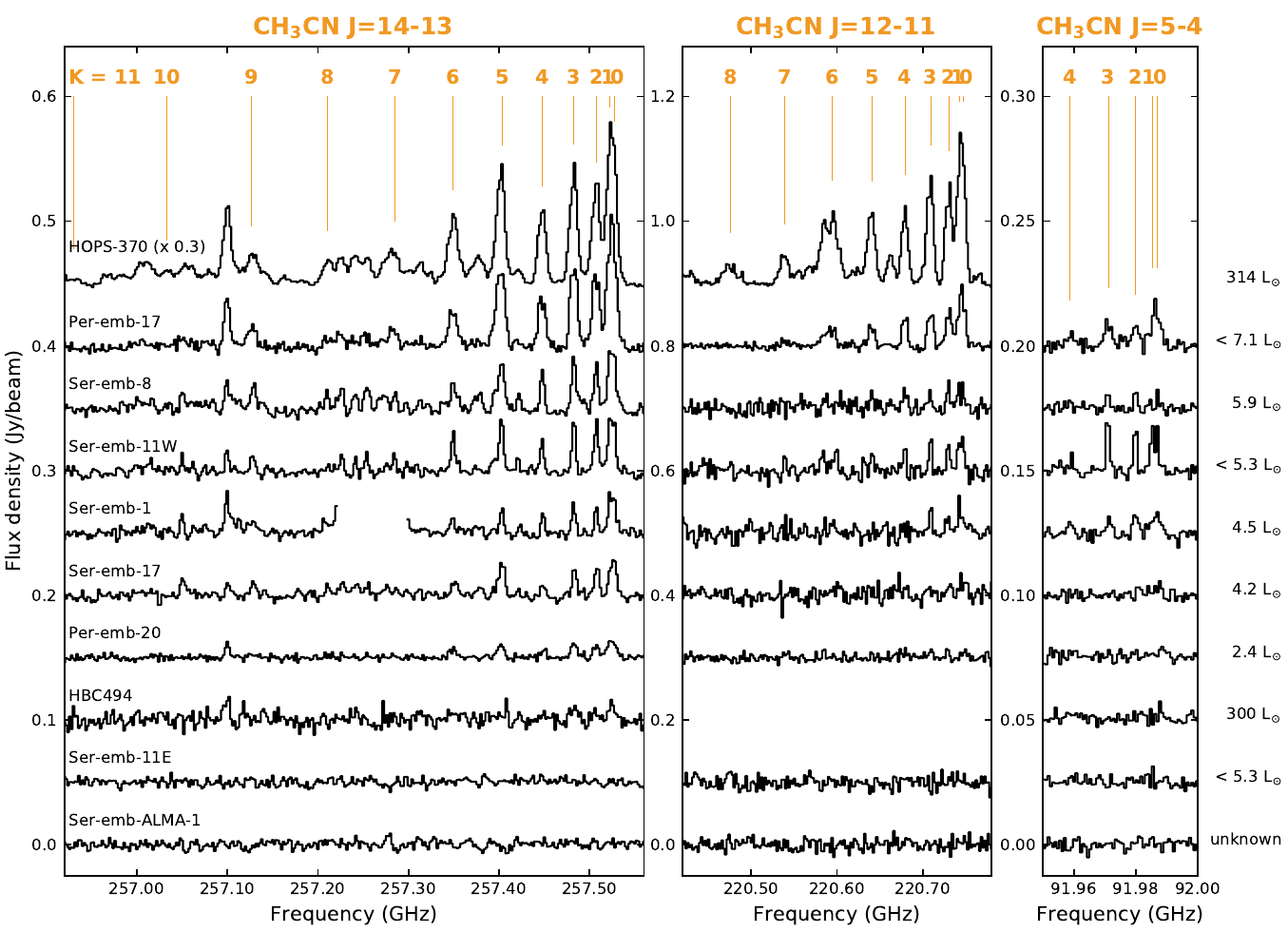}
\caption{Overview of CH$_3$CN $J_K=14_K-13_K$ (left panel), $J_K=12_K-11_K$ (middle panel), and $J_K=5_K-4_K$ (right panel) spectra at $\sim$2 km s$^{-1}$ resolution extracted toward the continuum peak for the sources in our sample. The vertical orange lines mark the different $K$-components in each ladder. Ser-emb-1 displays strong emission from $^{29}$SiO $J=6-5$ at 257.255202 GHz, and this region of the spectrum is omitted. The $14_5-13_5$ transition is blended with a CH$_3$OH line. The vertical scale is the same for all sources, except the $14_K-13_K$ spectrum towards HOPS-370 is scaled by a factor 0.3, but varies for the three different $J$-transitions. Bolometric luminosities are listed on the right, where a $<$ sign is used for binaries where the luminosity only has been measured for both components combined.}
\label{fig:CH3CN_observations}
\end{figure*}

For HOPS-370, we imaged the data without continuum subtraction and applied a statistical method as for example used by \citet{Jorgensen2016} for the line-rich spectrum of the protostellar binary IRAS 16293, to subtract the continuum from the spectra. This method entails the formation of a density distribution of flux values in the spectrum. If there would only be continuum emission, the density distribution would represent a symmetric Gaussian centered at the continuum level. In the case of continuum and line emission, the Gaussian will contain an exponential tail toward higher values. The continuum level can then determined by fitting a skewed Gaussian to the distribution. Given the large bandwidth of the NOEMA observations, we fit the density distribution per 1/8 of the bandwidth of each baseband ($\sim$1 GHz regions). The combination of high line density and intermediate spectral resolution results in double peaked distributions at 2 mm, where we take the peak at the lowest intensity as the continuum level. The line density is even higher at 1 mm, and the peak overpredicts the continuum level as it is located at intensity values $\gtrsim$7$\sigma$. We therefore estimate the continuum level as the intensity 3$\sigma$ above the intensity of the lowest bin. We then fit a first-order polynomial to the eight measurements for each baseband to take into account the continuum spectral index, and subtract this from the spectrum. The resulting spectra around the CH$_3$CN $J=14-13$ and $J=12-11$ ladders are displayed in Fig.~\ref{fig:CH3CN_observations}. 

\subsection{Rotation diagram analysis}\label{sec:methods_RD}

A rotation diagram analysis is a commonly used technique to derive column densities and excitation temperatures from molecular line emission \citep[e.g.,][]{Goldsmith1999}. In local thermodynamic equilibrium (LTE), the column density, $N$, and rotational temperature, $T_{\rm{rot}}$, can be calculated using the relation
\begin{equation} \label{eq:levelpopulations}
    \frac{N_{\rm{up}}}{g_{\rm{up}}} = \frac{N}{Q(T_{\rm{rot}})} \rm{e}^{(-E_{\rm{up}}/T_{\rm{rot}})}, 
\end{equation}
where $g_{\rm{up}}$ is the upper level degeneracy, $E_{\rm{up}}$ is the upper level energy, and $Q$ is the molecular partition function. $N_{\rm{up}}$ is the column density of the upper level, which can be determined from the observed line flux, $W = \int T_{\rm{mb}}dv$, assuming the emission is optically thin, using 
\begin{equation}
    N_{\rm{up}} = \frac{8\pi k \nu^2 W}{hc^3 A_{ul}}, 
\end{equation}
where $A_{ul}$ is the Einstein A coefficient and $\nu$ the frequency of the transition. The column density and rotational temperature can then be derived from the intercept and slope, respectively, of a linear fit to $\log(N_{\rm{up}}/g_{\rm{up}})$ versus $E_{\rm{up}}$. 

When the emission is not optically thin, an optical depth correction factor, $C_{\tau} = \tau/(1-e^{-\tau})$, must be applied to derive the true level populations:
\begin{equation} \label{eq:tau_correction}
    N_{\rm{up}} = N_{\rm{up}}^{\mathrm{obs}} C_{\tau}, 
\end{equation}
where the optical depth, $\tau$, depends on the upper level column density via
\begin{equation} \label{eq:tau}
    \tau = \frac{c^3 A_{ul} N_{\rm{up}}}{8 \pi \nu^3 \Delta v} (e^{h\nu/k T_{\rm{rot}}} - 1), 
\end{equation} 
where $\Delta v$ is the full width at half maximum (FWHM) of the line.

Even though some CH$_3$CN transitions appear optically thick in the rotation diagram, as will be described in Sect.~\ref{sec:RotationDiagrams}, the observed line fluxes cannot be reproduced with optically thick emission under the assumption that the unresolved emission fills the beam. \citet{Taquet2015} dealt with this by running a large grid of models with varying rotational temperature, column density and source size. On the other hand, \citet{Bergner2019} introduced a beam dilution factor to Eq.~\ref{eq:tau_correction}, based on an educated estimate of the emitting area, $\Omega_s$, such that
\begin{equation} \label{eq:tau_beam_correction}
    N_{\rm{up}} = N_{\rm{up}}^{\mathrm{obs}} C_{\tau} \frac{\Omega_b}{\Omega_s},
\end{equation}
where $\Omega_b$ is the solid angles of the beam. The beam and source solid angles can be expressed as $\Omega_b = \theta_{\rm{maj}}\theta_{\rm{min}}$ and $\Omega_s = \theta_s^2$, respectively, where $\theta_{maj}$ and $\theta_{min}$ are the major and minor axis of the beam, and $\theta_s$ is the emitting radius. Synthetic upper level populations with the total column density and rotational temperature as free parameters can then be generated by substituting Eqs.~\ref{eq:tau} and \ref{eq:tau_beam_correction} into Eq.~\ref{eq:levelpopulations}, and they are then fitted to the observed level populations (Eq.~\ref{eq:levelpopulations}) using the Markov Chain Monte Carlo (MCMC) package emcee \citep{Foreman-Mackey2013} to sample the posterior distribution \citep[e.g.,][]{Loomis2018}. Since we here want to investigate whether CH$_3$CN and CH$_3$OH have similar or different emitting areas toward eight different sources, we adopt the computationally more effective approach by \citet{Bergner2019}, but add the source solid angle as free parameter. Following \citet{Goldsmith1999} and \citet{Taquet2015}, we will use the term ``population diagram'' when corrections are applied for emitting area and optical depth and ``rotation diagram'' when these corrections are not made. Details about the line flux measurements and spectroscopic data are provided in Appendix~\ref{ap:LineFitting}.

\citet{Yang2021} concluded that LTE is a valid assumption for 1 mm CH$_3$OH emission toward the Perseus protostars in the PEACHES sample. Per-emb-20 is among the PEACHES sources with the lowest estimated gas densities ($\sim3\times10^{9}$ cm$^{-3}$) based on the continuum flux, and by far the weakest continuum source in our sample. LTE is also a good assumption for CH$_3$CN as the critical densities of the transitions are a few times $10^{6}$ cm$^{-3}$ for $14_K-13_K$ up to a few times $10^{7}$ cm$^{-3}$ for $5_K-4_K$ based on RADEX calculations (see also \citealt{Shirley2015}, Walls et al., under review).

\begin{figure*}
\centering
\includegraphics[width=\linewidth,trim={0.2cm 16.4cm 1cm 1.1cm},clip]{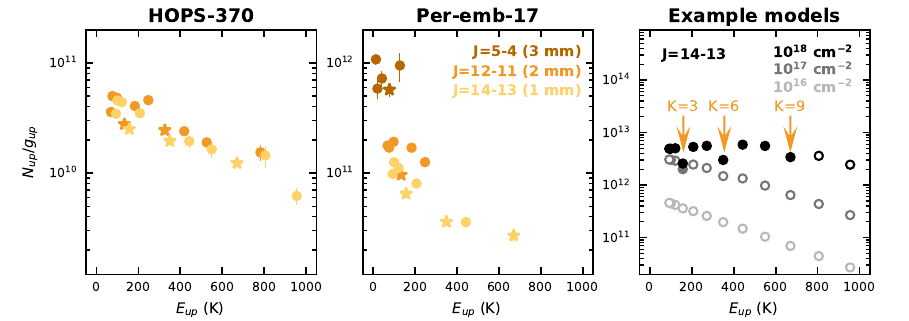}
\caption{Rotation diagrams for the CH$_3$CN $J=5-4$ (dark orange), $J=12-11$ (orange), and $J=14-13$ (yellow) ladders toward HOPS-370 and Per-emb-17. $K=3$, $K=6$ and $K=9$ transitions are shown as star symbols, the other transitions are shown as circles. The unresolved $12_K-11_K$ and $5_K-4_K$ fluxes have been scaled to the smaller beam size of the unresolved $14_K-13_K$ observations. The error bars correspond to 1$\sigma$, and are typically smaller than the symbols. Rotation diagrams for the other sources in the sample are shown in Fig.~\ref{fig:CH3CN_RD_total}. The right panel displays three LTE models for the $J=14-13$ ladder at 300 K with varying column densities (different shades of gray), highlighting the behavior of the $K=3$, $K=6$ and $K=9$ transitions when the emission becomes optically thick. Optically thin transitions in this panel are shown as open symbols, and optically thick transitions as filled symbols.} 
\label{fig:CH3CN_RD}
\end{figure*}


\section{Results and analysis}

\subsection{CH$_3$CN}\label{sec:CH3CN}

An overview of all CH$_3$CN observations is shown in Fig.~\ref{fig:CH3CN_observations}. $J_K = 14_K-13_K$ transitions are detected toward all sources except Ser-emb-11E and Ser-emb-ALMA-1. These latter two sources display very line-poor spectra and we exclude them from further analysis here. The $14_K-13_K$ lines are strongest toward the intermediate mass source HOPS-370, where $K$-components up to $K=11$ are detected. The brightest low-mass source is Per-emb-17, and all low-mass sources except Per-emb-20 and HBC494 display emission up to $14_9-13_9$. For Per-emb-20, the highest detected transition is $14_6-13_6$, and despite its high luminosity of $\sim$300 $L_\odot$, only $K=0-4$ are weakly detected toward HBC494. HBC494 has the strongest continuum peak in the sample after HOPS-370 (Table~\ref{tab:Observations}), so the line-poor spectrum is not due to a low mass of the circumstellar material, but may be related to the continuum optical depth. With the exception of HBC494, there seems to be a rough trend between line strength and bolometric luminosity, although the luminosities for Per-emb-17, Ser-emb-11E and Ser-emb-11W are the luminosities for both binary components combined. 

The $12_K-11_K$ transitions are only detected toward HOPS-370, Per-emb-17, Ser-emb-1, Ser-emb-8 and Ser-emb-11W. This can be attributed to both the $12_K-11_K$ transitions being intrinsically weaker than the $14_K-13_K$ transitions under given conditions, as well as the lower sensitivity of the $12_K-11_K$ observations. For HOPS-370, transitions up to $12_{10}-11_{10}$ are detected. For the brightest low-mass source the highest transition detected is $12_7-11_7$ (Ser-emb-11W), but the strong $12_9-11_9$ line is blended with $^{13}$CO. 

The $5_K-4_K$ transitions are much weaker than transitions from the other two ladders and are detected toward Per-emb-17, Ser-emb-1, Ser-emb-8, and Ser-emb-11W. The emission is particularly strong toward Ser-emb-11W. Except for Ser-emb-8, all $K$-components are detected. Only Ser-emb-8 shows signs of large scale emission being resolved out, especially for $5_0-4_0$ and $5_1-4_1$, as evidenced by large scale ripples with deep negative bowls in the velocity channels.

\begin{figure*}
\centering
\includegraphics[width=\linewidth,trim={0.2cm 16.4cm 1cm 1.1cm},clip]{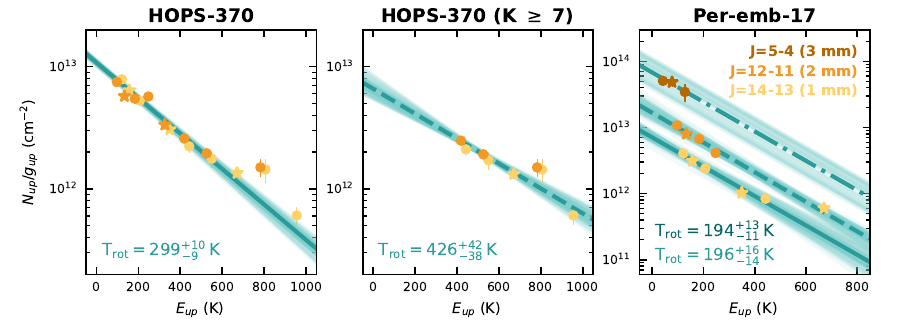}
\caption{Population diagram analysis for CH$_3$CN emission toward HOPS-370 and Per-emb-17. $K=3$, $K=6$ and $K=9$ transitions are shown as star symbols, the other transitions are shown as circles. The error bars correspond to 1$\sigma$, and are typically smaller than the symbols. For HOPS-370, the $J=12-11$ (orange), and $J=14-13$ (yellow) ladders are modeled together. In the left panel all transitions are included, while the middle panel shows the result for the optically thin transitions with $K \geq 7$. For Per-emb-17, the $J=5-4$ (dark orange), $J=12-11$ (orange), and $J=14-13$ (yellow) ladders are modeled individually (right panel). Draws from the fit posteriors are shown with light teal lines, with the 50th percentiles shown as thick teal lines. The corresponding rotational temperature is listed in the bottom left corner. For Per-emb-17, the top temperature is from the $J=12-11$ fit, the bottom from the $J=14-13$ fit. A solid line indicates that the rotational temperature, column density and source size were fitted, a dashed line indicates that the source size was fixed and a dash-dotted line indicates that the rotational temperature was fixed (see Table~\ref{tab:PD_CH3CN} for details and all fit results). Population diagrams for the other sources in the sample are shown in Figs.~\ref{fig:CH3CN_PD_14-13}-~\ref{fig:CH3CN_PD_5-4}.} 
\label{fig:CH3CN_PD}
\end{figure*}

\subsubsection{Rotation diagrams}\label{sec:RotationDiagrams}

Rotation diagrams including all detected CH$_3$CN transitions toward HOPS-370 and Per-emb-17 are presented in Fig.~\ref{fig:CH3CN_RD} and toward the other sources in Fig.~\ref{fig:CH3CN_RD_total}. The beam size of the observations increases with wavelength, but even in the smallest beam of the 1 mm observations ($J=14-13$) all emission is unresolved. Given the similar upper-level energies, we assume that the emission at 2 and 3 mm is also originating in a region smaller than the 1 mm beam. We therefore correct for the additional dilution within the larger beams at 2 and 3 mm by scaling the integrated intensities (in units of K km s$^{-1}$) at 2 and 3 mm by $\theta_{\rm{x}}^2/\theta_{\rm{1mm}}^2$, where $\theta_{\rm{x}}$ is the beam size at 2 or 3 mm. 

For optically thin emission, all transitions lie on a straight line with a negative slope. Because the $K=3,6,9$ transitions have double the statistical weight compared to the other $K$-transitions of a specific ladder but similar Einstein A coefficients, they can become optically thick while the other transitions remain optically thin. In this case, the optically thick line is weaker than expected for optically thin emission, which will result in the optically thick transition to end up below the linear trend in the rotation diagram (see Fig.~\ref{fig:CH3CN_RD}, right panel). If the column is high enough for all $K$-components to be optically thick, they will form a straight line with a slight positive slope with the $K=3,6,9$ transitions below the trend (see Fig.~\ref{fig:CH3CN_RD}, right panel). Finally, for optically thick emission, the $12_K-11_K$ transitions will be located slightly above the $14_K-13_K$ transitions, and the $5_K-4_K$ transitions will be even higher up for a given column density and temperature. 

With the above in mind, it is clear from Figs.~\ref{fig:CH3CN_RD} and \ref{fig:CH3CN_RD_total} that the $J=14-13$ ladder is only fully optically thin toward HBC494. For all other sources, at least the $14_3-13_3$ transition is optically thick. For Ser-emb-1, Ser-emb-8 and Ser-emb-17, the $K=7-9$ transitions appear overly strong compared to the lower $K$-transitions, suggesting that there may be two temperature components.

The $J=12-11$ ladder is detected toward five of the eight sources in the sample (Fig.~\ref{fig:CH3CN_RD_total}), but only for HOPS-370 are the $14_K-13_K$ and $12_K-11_K$ transitions showing the same behavior. For the other sources with detections, the $12_K-11_K$ transitions are typically higher up in the rotation diagram than $14_K-13_K$ transitions with similar $E_{\rm{up}}$. This suggest that they are either (more) optically thick, and/or trace a larger column of material than $14_K-13_K$. 

Ser-emb-1 is the only source where the $12_K-11_K$ transitions appear optically thin, although it is difficult to determine with no detections of $14_4-13_4$ and $14_5-13_5$ and a tentative detection of $14_6-13_6$. For Per-emb-17, Ser-emb-8 and Ser-emb-11W, all observed $12_K-11_K$ transitions appear at least marginally optically thick with $12_3-11_3$ and $12_6-11_6$ (if detected) clearly below the linear trend of the entire ladder and a nearly horizontal slope. The $12_K-11_K$ transitions thus appear more optically thick than the $14_K-13_K$ transitions toward these three sources. 

\begin{deluxetable*}{@{\extracolsep{4pt}}lccccccccccc}
\tablecaption{Rotational temperatures, column densities and source sizes for CH$_3$CN. \label{tab:PD_CH3CN}}
\tablewidth{0pt}
\addtolength{\tabcolsep}{-3pt} 
\tabletypesize{\scriptsize}
\tablehead{
\colhead{} & \multicolumn{3}{c}{$J=14-13$ (1 mm)} & \multicolumn{3}{c}{$J=12-11$ (2 mm)} & \multicolumn{3}{c}{$J=5-4$ (3 mm)} \\
\cline{2-4} \cline{5-7} \cline{8-10}
\colhead{Source} \vspace{-0.3cm} & \colhead{$T_{\rm{ex}}$} & \colhead{$N$} & \colhead{$\theta_s$} & \colhead{$T_{\rm{ex}}$} & \colhead{$N$} & \colhead{$\theta_s$} & \colhead{$T_{\rm{ex}}$} & \colhead{$N$} & \colhead{$\theta_s$} \\
\colhead{} & \colhead{(K)} & \colhead{(cm$^{-2}$)} & \colhead{(arcsec)} & \colhead{(K)} & \colhead{(cm$^{-2}$)} & \colhead{(arcsec)} & \colhead{(K)} & \colhead{(cm$^{-2}$)} & \colhead{(arcsec)} 
} 
\startdata 
HBC494      & $\lesssim$ 388 & $\lesssim$ $4.0\times10^{15}$ & [0.122]$^a$ & 
              \nodata & \nodata & \nodata & 
              [388] & $<$ $1.1\times10^{17}$ & [0.122]$^a$ \\
HOPS-370    & $278^{+14}_{-13}$ & ($1.4^{+0.3}_{-0.3}$)$\times10^{17}$ & $0.151^{+0.004}_{-0.004}$ & 
              $273^{+19}_{-17}$ & ($1.9^{+0.4}_{-0.4}$)$\times10^{17}$ & $0.132^{+0.003}_{-0.003}$ & 
              \nodata & \nodata & \nodata \\
Per-emb-17  & $195^{+15}_{-14}$ & ($4.4^{+1.1}_{-1.0}$)$\times10^{16}$ & $0.089^{+0.003}_{-0.003}$ & 
              $194^{+13}_{-11}$ & ($1.0^{+0.2}_{-0.2}$)$\times10^{17}$ & [0.089] & 
              [195] & ($4.2^{+1.7}_{-1.4}$)$\times10^{17}$ & $0.080^{+0.006}_{-0.005}$ \\
Per-emb-20  & $343^{+99}_{-93}$ & ($1.1^{+1.3}_{-0.7}$)$\times10^{17}$ & $0.030^{+0.006}_{-0.004}$ & 
              [343] & $<$ $9.7\times10^{16}$ & [0.030] & 
              [343] & $<$ $1.0\times10^{18}$ & [0.030] \\
Ser-emb-1   & $388^{+75}_{-86}$ & ($2.1^{+1.5}_{-1.1}$)$\times10^{17}$ & $0.034^{+0.005}_{-0.003}$ & 
              $-$ & $-$ & $-$ & 
              [388] & ($4.2^{+2.3}_{-1.7}$)$\times10^{18}$ & $0.050^{+0.004}_{-0.003}$ \\
Ser-emb-8   & $358^{+54}_{-46}$ & ($1.6^{+0.8}_{-0.5}$)$\times10^{17}$ & $0.049^{+0.004}_{-0.004}$ &
              $355^{+49}_{-41}$ & ($5.3^{+2.5}_{-2.0}$)$\times10^{17}$ & [0.049] & 
              [358] & ($2.4^{+0.8}_{-0.9}$)$\times10^{18}$ & [0.049] \\
Ser-emb-11W  & $309^{+49}_{-41}$ & ($7.5^{+3.3}_{-2.4}$)$\times10^{16}$ & $0.056^{+0.004}_{-0.004}$ & 
               $343^{+33}_{-31}$ & ($4.4^{+1.1}_{-0.9}$)$\times10^{17}$ & [0.056] & 
               [309] & ($1.8^{+1.2}_{-0.7}$)$\times10^{17}$ & $0.15^{+0.03}_{-0.02}$ \\
Ser-emb-17  & $347^{+89}_{-79}$ & ($1.5^{+1.4}_{-0.8}$)$\times10^{17}$ & $0.033^{+0.006}_{-0.004}$ &
              [347] & $<$ $1.3\times10^{17}$ & [0.033] &
              [347] & $<$ $6.9\times10^{17}$ & [0.033] \\
\enddata
\vspace{0.1cm}
\textbf{Notes.} Rotational temperature,  column density and source size determined from population diagram analysis. Listed values and uncertainties correspond to the 50th, and 16th and 84th percentiles, respectively, of the posterior distributions. For the column densities, a flux calibration uncertainty (20\% at 1 and 2 mm, 10\% at 3 mm) has been added to the fit uncertainty through a root sum square. For sources with no $5_K-4_K$ and/or $12_K-11_K$ detections upper limits are calculated using the rotational temperature and source size derived from the $14_K-13_K$ transitions. Values listed between square brackets are kept fixed at the value derived from the $14_K-13_K$ transitions, unless noted otherwise. Three dots (\nodata) indicate that a frequency range has not been observed, and a dash ($-$) indicates that not enough transitions have been detected at high enough signal to noise to obtain a reliable fit. For HOPS-370, the combined fit to the $14_K-13_K$ and $12_K-11_K$ transitions combined results in $299\pm10$ K and $(1.6\pm0.3)\times10^{17}$ cm$^{-2}$.
\vspace{-0.3cm}
\tablenotetext{a}{Source size kept fixed at the deconvolved continuum size.}
\end{deluxetable*}

The $5_K-4_K$ transitions are located above the $12_K-11_K$ and $14_K-13_K$ transitions in the rotation diagram for all sources with detections (Figs.~\ref{fig:CH3CN_RD} and \ref{fig:CH3CN_RD_total}). Again, suggesting that they are either optically thick, and/or trace a larger column of material. While the $K=0-2$ components get optically thick at low temperatures ($\lesssim$ 30 K) for column densities a few times $10^{15}$ cm$^{-2}$, the $K=3$ and $K=4$ components become most easily optically thick at temperatures of $30-40$ K, which then requires column densities of $\sim$10$^{16}$ cm$^{-2}$ and $\sim$10$^{17}$ cm$^{-2}$, resp (Fig.~\ref{fig:tau_CH3CN5-4}). In addition, for the $5_4-4_4$ transition to become half the strength of the $5_2-4_2$ transition (as observed; Fig.~\ref{fig:CH3CN_observations}), column densities $\gtrsim10^{17}$ cm$^{-2}$ are required (Fig.~\ref{fig:tau_CH3CN5-4}). Such high columns of cold material are unlikely as CH$_3$CN column densities of a few times $10^{11}-10^{12}$ cm$^{-2}$ are reported toward dark clouds and prestellar cores \citep{Minh1993,Vastel2019,Megias2023}. This suggests that the observed $5_K-4_K$ emission is not simply tracing optically thick cold emission from the outer envelope. Instead, the $5_K-4_K$ transitions at 90 GHz are likely to trace a larger column of material than the $14_K-13_K$ and $12_K-11_K$ transitions because of the lower dust opacity at these longer wavelengths (3 mm).

\subsubsection{Population diagram analysis}

Due to the combination of optically thin and thick lines, we can constrain the rotational temperature, column density and emitting area from the $14_K-13_K$ transitions as described in Sect.~\ref{sec:methods_RD}. Since the $14_0-13_0$ and $14_1-13_1$ transitions are blended we exclude them from the population diagram analysis. The population diagrams for HOPS-370 and Per-emb-17 are presented in Fig.~\ref{fig:CH3CN_PD}, and for the other sources in Fig.~\ref{fig:CH3CN_PD_14-13}. The best fit parameters, defined as the 50th percentile of the posterior distributions, are listed in Table~\ref{tab:PD_CH3CN}.

\begin{figure}
\centering
\includegraphics[trim={0cm 13.7cm 0cm 1.5cm},clip]{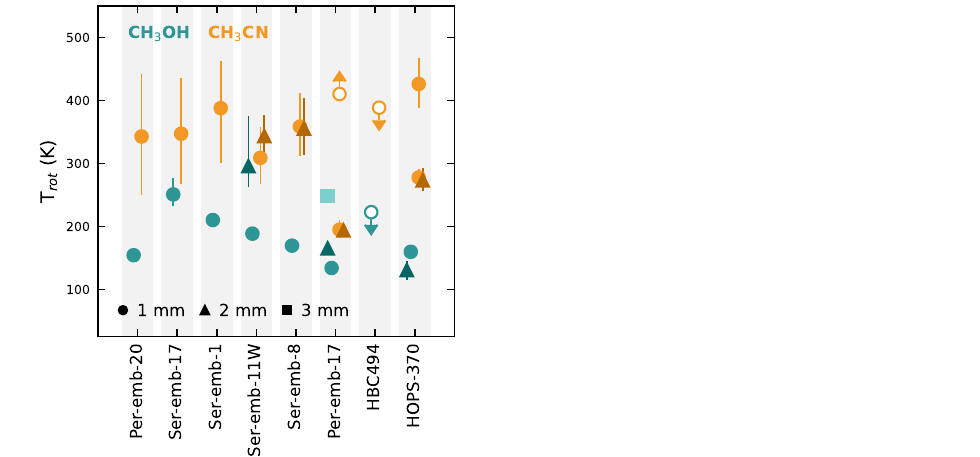}
\caption{Overview of rotational temperatures derived for CH$_3$CN (orange) and CH$_3$OH (teal), with sources ordered by their bolometric luminosity. Temperatures derived from the 1 mm data ($J_K=14_K-13_K$ for CH$_3$CN) are shown as circles, from the 2 mm data ($J_K=12_K-11_K$ for CH$_3$CN) as triangles, and from the 3 mm data as squares. The highest orange circles for Per-emb-17 and HOPS-370 are derived from the high-$K$ transitions only. Upper and lower limits are indicated with open symbols with an arrow pointing up or down, respectively.}
\label{fig:Trot_source}
\end{figure}

The derived rotational temperatures from the $14_K-13_K$ transition range between 195-390 K (see also Fig.~\ref{fig:Trot_source}). While reasonable fits can be obtained to all $14_K-13_K$ transitions for most sources, the highest $K$-transitions toward HOPS-370 and Per-emb-17 can not be reproduced with a single temperature component. For HOPS-370, the transitions with $K > 6$ are optically thin ($\tau < 0.35$). Fitting these transitions separately (with a fixed source size) results in a rotational temperature of $426^{+42}_{-38}$ K (Fig.~\ref{fig:CH3CN_PD}). For Per-emb-17, fewer high $K$ transitions are detected and $14_6-13_6$ is marginally optically thick ($\tau = 0.6$). The rotational temperature from a fit to the high $K$ transitions alone is therefore not well constrained, but suggests $T_{\rm{rot}} \gtrsim 400$ K. For Ser-emb-1, Ser-emb8, and Ser-emb-17, there seems to be a change in slope of the population diagram for $K > 6$, but given the small number of transitions and the emission being unresolved, fits to the high $K$ components are not well constrained. 

As expected from the rotation diagrams (Sect.~\ref{sec:RotationDiagrams}), only for HOPS-370 can the $14_K-13_K$ and $12_K-11_K$ transitions be reproduced by a single model. We therefore analyze the $12_K-11_K$ transitions separately. However, with fewer lines detected than for the $J=14-13$ ladder and potentially fully optically thick emission, we cannot constrain the rotational temperature, column density and source size simultaneously from the $J=12-11$ ladder. We therefore fix the source size to that derived from the $14_K-13_K$ transitions. The resulting population diagram for Per-emb-17 is presented in Fig.~\ref{fig:CH3CN_PD}, and for the other sources in Fig.~\ref{fig:CH3CN_PD_12-11}. All fit parameters are listed in Table~\ref{tab:PD_CH3CN}.

While the resulting rotational temperatures are similar to the ones derived from the $14_K-13_K$ transitions, the column densities are a factor of a few higher. Probing higher column densities at 2 mm could either be a reflection of the $12_K-11_K$ transitions becoming optically thick at slightly higher columns than the $14_K-13_K$ transitions due to lower Einstein A coefficients, or because the dust opacity is slightly lower at 2 mm compared to 1 mm allowing more material to be observed. The more optically thick nature of the $12_K-11_K$ transitions in the rotation diagram, as particularly evident for Ser-emb-11W, favors the latter explanation. 

\begin{figure*}
\centering
\includegraphics[trim={0cm 15.5cm 0cm 1.5cm},clip]{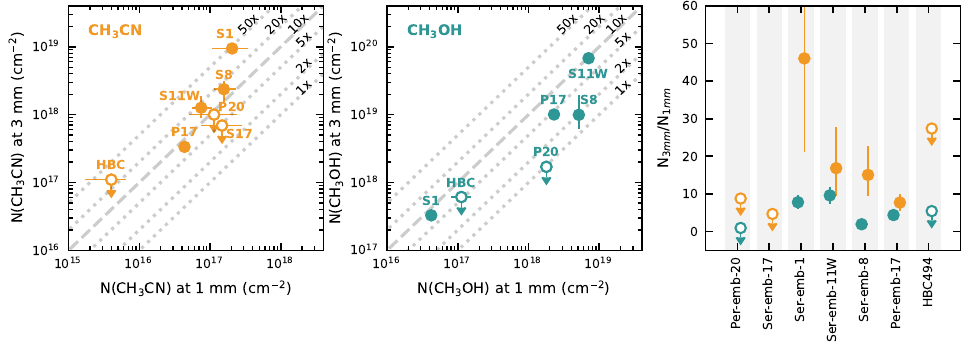}
\caption{Comparison of column densities derived at 1 mm ($J_K=14_K-13_K$ for CH$_3$CN) and 3 mm ($J_K=5_K-4_K$ for CH$_3$CN). The left and middle panel show the 3 mm column versus the 1 mm column for CH$_3$CN and CH$_3$OH, respectively. Sources are identified by the first letter and the number in their names, except HBC494 which is abbreviated to HBC. The dotted and dashed gray lines mark increases in the 3 mm column of 1, 2, 5, 10 (dashed), 20 and 50 times compared to the 1 mm column, going from bottom right to top left. The right panel shows the ratio between the column density at 3 mm and 1 mm for CH$_3$CN (orange) and CH$_3$OH (teal) per source, with sources ordered by bolometric luminosity. For sources where a source size could be constrained for the 3 mm CH$_3$CN emission the column densities are scaled to the 1 mm source size. No 3 mm observations were performed toward HOPS-370. Upper and lower limits are indicated with open symbols with arrows.  }
\label{fig:N_1mm-3mm}
\end{figure*}

The $J=5-4$ ladder only has three unblended transitions ($K=2-4$) so we cannot constrain rotational temperature, column density and source size simultaneously. We fix the rotational temperature to the value derived from the $14_K-13_K$ transitions, because the rotational temperature cannot be constrained without knowledge of the emitting area for optically thick emission, and if we are indeed tracing more material at 3 mm due to the lower dust opacity, the emitting areas at 1 and 3 mm will not be the same. The resulting population diagram for Per-emb-17 is presented in Fig.~\ref{fig:CH3CN_PD}, and for the other sources in Fig.~\ref{fig:CH3CN_PD_5-4}. All fit parameters are listed in Table~\ref{tab:PD_CH3CN}. For the sources with a detection of the $5_K-4_K$ transitions, the column density is a factor 8--46 higher than derived from the $14_K-13_K$ transitions when scaling to the same emitting area (Fig.~\ref{fig:N_1mm-3mm}). The smallest increase is derived for Per-emb-17 and the largest increase toward Ser-emb-1. The upper limits based on the $5_K-4_K$ nondetections toward HBC494, Per-emb-20 and Ser-emb-17 are consistent with column density increases at 3 mm up to a factor 27, 8.9 and 4.7, respectively. The derived emitting area at 3 mm is only a factor 1.5 and 2.7 larger for Ser-emb-1 and Ser-emb-11W, respectively, while it is a factor 0.9 smaller for Per-emb-17. The source size was fixed in the fit for Ser-emb-8 because the $5_4-4_4$ transition was not detected. The increases in column density at 3 mm are thus not due to a smaller emitting area retrieved from the fit. Instead, the larger column densities and emitting areas suggest that we are see more material closer to the star at 3 mm than at 1 mm. 

Overall, the CH$_3$CN emission shows rotational temperatures higher than $\sim$300 K, based on the $J=14-13$ ladder. In addition, we may be probing larger columns with the $5_K-4_K$ transitions at 90 GHz, and maybe already with the $12_K-11_K$ transitions at 220 GHz, which would be consistent with lower dust opacities at lower frequencies.

\subsection{CH$_3^{13}$CN} \label{sec:CH313CN}

The CH$_3^{13}$CN transitions are located close to the corresponding main isotopologue transitions in frequency space. We do not detect CH$_3^{13}$CN toward any of the sources in the sample, but we are able to put constraints on the excitation temperature of a potentially second temperature component for Ser-emb-1, Ser-emb-8 and Ser-emb-17 using the non-detection of CH$_3^{13}$CN $J_K=14_K-13_K$. For increasing temperatures, the CH$_3^{13}$CN column density upper limit derived from the 3$\sigma$ noise level increases, while at the same time a lower column is required to reproduce the high-$K$ transitions of CH$_3$CN. This means that the combination of observational constraints places a lower limit on the excitation temperature, as for too low temperatures the column density needed to explain the high-$K$ CH$_3$CN emission should have resulted in a detection of CH$_3^{13}$CN. 

Our approach is as follows. We model the $14_6-13_6$ to $14_9-13_9$ transitions using a fixed temperature. Using this temperature and the best fit column density (or the lowest column that results in optically thick emission), we calculate the expected CH$_3^{13}$CN column assuming a $^{12}$C/$^{13}$C ratio of 68 \citep{Milam2005}. If the expected CH$_3^{13}$CN column exceeds the derived upper limit, this process is repeated with a 10 K higher excitation temperature. To be in agreement with the CH$_3^{13}$CN nondetection, the excitation temperatures then need to be $\gtrsim$250, $\gtrsim$300, and $\gtrsim$290 K, respectively, for Ser-emb-1, Ser-emb-8 and Ser-emb-17. This confirms the presence of $\sim$300 K gas toward these sources.


\subsection{CH$_3$OH}\label{sec:CH3OH}

\begin{figure*}
\centering
\includegraphics[width=\linewidth,trim={0.2cm 17.3cm 1cm 1.1cm},clip]{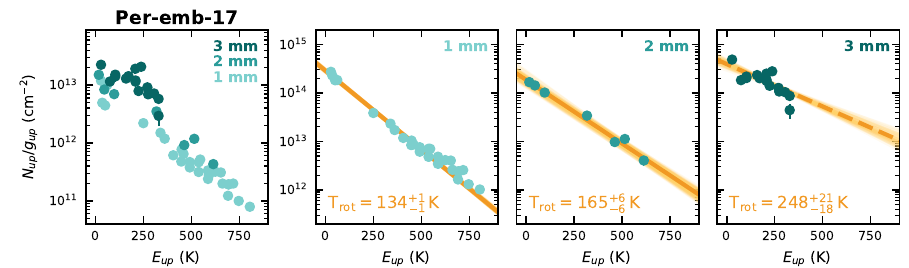}
\caption{Rotation diagrams for CH$_3$OH at 1 mm (light teal), 2 mm (teal) and 3 mm (dark teal) toward Per-emb-17 (left panel) and results from a population diagram analysis of the observations at different wavelengths (three right most panels). The error bars correspond to 1$\sigma$, and are typically smaller than the symbols. Draws from the fit posteriors are shown with light orange lines, with the 50th percentiles shown as thick orange lines. The corresponding rotational temperature is listed in the bottom left corner. A solid line indicates that the rotational temperature, column density and source size were fitted, while a dashed line indicates that the source size was fixed (see Table~\ref{tab:PD_CH3OH} for details and all fit results). Rotation diagrams for all sources are presented in Fig.~\ref{fig:CH3OH_RD_total}, and results from the population diagram analysis are shown in Figs.~\ref{fig:CH3OH_PD_1mm}-~\ref{fig:CH3OH_PD_3mm}.}
\label{fig:CH3OH_RD}
\end{figure*}

The 1 mm setting covers the most CH$_3$OH transitions, and for all sources except HBC494, transitions with upper level energies ranging between $\sim$30 and  $\sim$600 K are detected (Fig.~\ref{fig:CH3OH_RD} and Table~\ref{tab:CH3OH_parameters}). The highest upper level energy detected toward HBC494 is 73 K, while toward HOPS-370, Per-emb-17, Ser-emb-8 and Ser-emb-11W transitions with upper level energies up to $\sim$800 K are detected. The 2 mm setting covers the least CH$_3$OH transitions. The most lines are detected toward HOPS-370, Per-emb-17 and Ser-emb-11W, with upper level energies ranging between $\sim$15--600 K. No transitions are detected at 2 mm toward Ser-emb-1 and Ser-emb-8, and no 2 mm observations were carried out toward HBC494. The transitions detected in the 3 mm setting span a smaller range of upper level energies ($\sim$30--330 K), but transitions with lower Einstein A coefficients are detected (down to $8\times10^{-7}$ s$^{-1}$ versus $1\times10^{-6}$ s$^{-1}$ at 1 mm). No transitions are detected at 3 mm toward HBC494 and Per-emb-20, and no 3 mm observations were carried out toward HOPS-370.

Rotation diagrams for CH$_3$OH are presented in Fig.~\ref{fig:CH3OH_RD} for Per-emb-17 and in Fig.~\ref{fig:CH3OH_RD_total} for all sources in the sample. Similar as seen for CH$_3$CN, transitions at longer wavelengths are typically located higher up in the rotation diagram. Results of the population diagram analysis are presented in Fig.~\ref{fig:CH3OH_RD} for Per-emb-17 and in Figs.~\ref{fig:CH3OH_PD_1mm}-\ref{fig:CH3OH_PD_3mm} for the other sources. All results are listed in Table~\ref{tab:PD_CH3OH}. The 1 mm transitions are generally at least partially optically thick, except for Ser-emb-1 where the emission appears optically thin. The derived rotational temperatures at 1 mm range between 134 $\pm$ 1 K (Per-emb-17) and $251^{25}_{-19}$ K (Ser-emb-17) (see also Fig.~\ref{fig:Trot_source}). In contrast to CH$_3$CN, there is no evidence for two temperature components. The rotational temperature derived for HBC494 has a large uncertainty because only a few low-energy transitions are detected. The non-detection of the $17_{3,14}-17_{2,15}$ transition at 248.282424 GHz ($E_{\rm{up}}$= 405 K) provides the strongest constraint on the temperature and restricts it to values below $\sim$223~K.

For HOPS-370, the 1 and 2 mm data points are again in good agreement. For L1455-IRS1, the 2 mm data points are visibly higher up in the rotation diagram and a fit to the 2 mm data results in a slightly higher temperature (165 $\pm$ 6 K versus 134 $\pm$ 1 K), and a similar column density ($2.3\pm0.5\times10^{18}$ cm$^{-2}$ versus $2.2\pm0.5\times10^{18}$) for a larger emitting area ($0.139\pm0.006$ versus $0.117\pm0.001$). Due to the low number of detections, excitation temperature, column density and source size cannot be constrained simultaneously for the other sources at 2 mm. For Ser-emb-11W, fixing the emitting area to that derived at 1 mm also results in a higher temperature and factor $\sim$2 higher column density. For Per-emb-20 and Ser-emb-1, the 2 mm results are consistent with the conditions derived at 1 mm, while for Ser-emb-8 the upper limit is an order of magnitude lower than the 1 mm column density. For Ser-emb-17, the large scatter between the two detections does not allow for a firm conclusion to be drawn.

For Per-emb-17, Ser-emb-1 and Ser-emb-11W, the 3 mm data points are clearly higher up in the rotation diagram compared to the transitions at 1 and 2 mm. For Per-emb-17, the transitions appear optically thin ($\tau < 0.3$ for all but one transition), so we fix the source size to the value derived at 1 mm. This then results in a higher temperature ($248^{+21}_{-18}$ K) and higher column (1.0 $\pm$ 0.1 $\times10^{19}$ cm$^{-2}$) compared to both the 1 mm (134 $\pm$ 1 K and 2.3 $\pm$ 0.5 $\times 10^{18}$ cm$^{-2}$) and the 2 mm data (165 $\pm$ 6 K and 2.2 $\pm$ 0.5 $\times 10^{18}$ cm$^{-2}$) (see Figs.~\ref{fig:N_1mm-3mm} and \ref{fig:CH3OH_RD}).  

Again, due to larger scatter (Ser-emb-8 and Ser-emb-11W) or few detections (Ser-emb-1 and Ser-emb-17) we cannot reliably fit the 3 mm data for the other sources. However, assuming a rotational temperature and source size equal to that obtained at 1 mm results in 3 mm columns a factor $\sim$1.9--9.6 higher than at 1 mm (see also Fig.~\ref{fig:N_1mm-3mm}). For Ser-emb-17, the single detected transition cannot be reproduced with the rotational temperature and source size derived at 1 mm. No 3 mm transitions are detected toward HBC494 and Per-emb-20, but column density upper limits allow for columns a factor 1.0--5.4 higher at 3 mm (Table~\ref{tab:PD_CH3OH} and Fig.~\ref{fig:N_1mm-3mm}). 

Taken together, the CH$_3$OH rotational temperature shows little variation among the sources in our sample (135--250 K) based on the transitions detected at 1 mm. For Per-emb-17, both the rotational temperature and the column density increase for transitions detected at longer wavelengths. The three other sources with multiple detections at 3 mm (Ser-emb-1, Ser-emb-8 and Ser-emb-11W) also show column density increases of a factor $\sim$2--10 at 3 mm. This confirms the CH$_3$CN results that we are seeing hotter gas closer to the protostar at 3 mm.

\begin{deluxetable*}{@{\extracolsep{4pt}}lccccccccccc}
\tablecaption{Rotational temperatures, column densities and source sizes for CH$_3$OH. \label{tab:PD_CH3OH}}
\tablewidth{0pt}
\addtolength{\tabcolsep}{-3pt} 
\tabletypesize{\scriptsize}
\tablehead{
\colhead{} & \multicolumn{3}{c}{1 mm} & \multicolumn{3}{c}{2 mm} & \multicolumn{3}{c}{3 mm} \\
\cline{2-4} \cline{5-7} \cline{8-10}
\colhead{Source} \vspace{-0.3cm} & \colhead{$T_{\rm{ex}}$} & \colhead{$N$} & \colhead{$\theta_s$} & \colhead{$T_{\rm{ex}}$} & \colhead{$N$} & \colhead{$\theta_s$} & \colhead{$T_{\rm{ex}}$} & \colhead{$N$} & \colhead{$\theta_s$} \\
\colhead{} & \colhead{(K)} & \colhead{(cm$^{-2}$)} & \colhead{(arcsec)} & \colhead{(K)} & \colhead{(cm$^{-2}$)} & \colhead{(arcsec)} & \colhead{(K)} & \colhead{(cm$^{-2}$)} & \colhead{(arcsec)} 
} 
\startdata 
HBC494      & $\lesssim$ 223 & $\lesssim$ $1.1\times10^{17}$ & [0.122]$^a$
            & \nodata & \nodata & \nodata
            & [223] & $<$ $6.1\times10^{17}$ & [0.122]$^a$ \\
HOPS-370    & $160^{+3}_{-4}$ & $(6.6^{+1.0}_{-1.0})\times10^{18}$ & $0.158^{+0.001}_{-0.001}$
            & $131^{+15}_{-16}$ & $(1.3^{+0.4}_{-0.3})\times10^{19}$ & $0.138^{+0.010}_{-0.007}$
            & \nodata & \nodata & \nodata \\
Per-emb-17  & $134^{+1}_{-1}$ & $(2.3^{+0.5}_{-0.5})\times10^{18}$ & $0.117^{+0.001}_{-0.001}$
            & $165^{+6}_{-6}$ & $(2.2^{+0.5}_{-0.5})\times10^{18}$ & $0.139^{+0.006}_{-0.005}$
            & $248^{+21}_{-18}$ & $(1.0^{+0.1}_{-0.1})\times10^{19}$ & [0.117] \\
Per-emb-20  & $154^{+7}_{-6}$ & $(1.8^{+0.4}_{-0.4})\times10^{18}$ & $0.045^{+0.002}_{-0.002}$
            & [154] & $\sim 2.0\times10^{18}$ & [0.045]
            & [154] & $<$ $1.7\times10^{18}$ & [0.045] \\
Ser-emb-1   & $210^{+12}_{-11}$ & $(4.2^{+0.9}_{-0.9})\times10^{16}$ & [0.225]$^a$
            & [210] & $<$ $6.3\times10^{16}$ & [0.225]$^a$
            & [210] & $(3.3^{+0.7}_{-0.7})\times10^{17}$ & [0.225]$^a$ \\
Ser-emb-8   & $169^{+7}_{-7}$ & $(5.1^{+1.2}_{-1.2})\times10^{18}$ & $0.062^{+0.002}_{-0.002}$
            & [169] & $<$ $5.6\times10^{17}$ & [0.062]
            & [169] & $(9.9^{+9.4}_{-3.9})\times10^{18}$ & [0.062] \\
Ser-emb-11W & $188^{+9}_{-9}$ & $(7.1^{+1.5}_{-1.5})\times10^{18}$ & $0.063^{+0.002}_{-0.002}$
            & $296^{+80}_{-32}$ & $(1.4^{+0.6}_{-0.4})\times10^{19}$ & [0.063]
            & [296]$^b$ & $(6.8^{+1.5}_{-1.2})\times10^{19}$ & [0.063] \\
Ser-emb-17  & $251^{+25}_{-19}$ & $(1.6^{+0.4}_{-0.4})\times10^{19}$ & $0.038^{+0.002}_{-0.002}$
            & $-$ & $-$ & $-$
            & $-$ & $-$ & $-$
\enddata
\vspace{0.1cm}
\textbf{Notes.} Rotational temperatures, column densities and source sizes obtained from population diagram analysis. Listed values and uncertainties correspond to the 50th, and 16th and 84th percentiles, respectively, of the posterior distributions. For the column densities, a flux calibration uncertainty (20\% at 1 and 2 mm, 10\% at 3 mm) has been added to the fit uncertainty through a root sum square. Values listed between square brackets are kept fixed at the value derived from the 1 mm transitions, unless noted otherwise. For sources with no detections at 2 and/or 3 mm, upper limits are calculated using the rotational temperature and source size derived from the 1 mm transitions. A dash ($-$) indicates that no column density could be determined. Three dots (\nodata) indicate that a frequency range has not been observed. For HOPS-370, the combined fit to the 1 mm and 2 mm data combined results in $167\pm2$ K and $(7.2\pm0.1)\times10^{18}$ cm$^{-2}$.
\vspace{-0.2cm}
\tablenotetext{a}{Source size fixed at deconvolved continuum size.}
\tablenotetext{b}{Temperature fixed at the value derived at 2 mm.}
\end{deluxetable*}


\section{Discussion}\label{sec:Discussion}

For all eight sources in our sample with detections of CH$_3$CN and CH$_3$OH, the rotational temperature for CH$_3$CN is found to be higher than for CH$_3$OH (Fig.~\ref{fig:Trot_source}). In addition, the CH$_3$CN temperatures are generally higher, with rotational temperatures at 1 mm for CH$_3$CN ranging between $\sim$280--425 K (except for Per-emb-17, where $T_{\rm{rot}}$ = 195 K, but with a strong indication of a second warmer component) and for CH$_3$OH between $\sim$135--250 K. In Sect.~\ref{sec:spatialdistribution}, we will show that this difference in rotational temperatures is indicative of different spatial distributions, with CH$_3$CN enhanced in hot gas. We will discuss potential origins of this enhancement in Sect.~\ref{sec:enhancement}, arguing that it may be the result of carbon-grain sublimation. Finally, we will discuss in Sect.~\ref{sec:othersources} that carbon-grain sublimation may then be common around low-mass protostars.

\subsection{Spatial distribution of CH$_3$CN and CH$_3$OH} \label{sec:spatialdistribution}

There is no correlation between rotational temperature and luminosity. While HOPS-370 is more than 50 times more luminous than the low-mass sources (except HBC494), the CH$_3$CN rotational temperature is only marginally higher and the CH$_3$OH rotational temperature is even amongst the lowest in the sample. This hints that CH$_3$CN emission is more effected by the source luminosity, and therefore potentially the presence of hot circumstellar material, than CH$_3$OH. Due to the weak emission toward HBC494, it is not possible to tell whether this source shows the same trend as HOPS-370 which has similar luminosity. 

Since both molecules are observed simultaneously and transitions with a similar range of upper-level energies are covered, the observed difference in rotational temperature is not due to differences in instrumental setup. Instead, it points to a different spatial distribution of the molecules, with CH$_3$CN tracing hotter gas than CH$_3$OH. This could mean that CH$_3$CN has a more compact distribution than CH$_3$OH or that the CH$_3$CN abundance is enhanced in hot gas compared to cooler gas, while the CH$_3$OH abundance is more constant with radius or decreases at higher temperature. Difference in spatial extent or the radial intensity profile are not directly visible in our spatially unresolved observations, but we can assess the size of the emitting region using both optically thick and thin transitions, and the relative contributions from warm versus hot gas by comparing emission at 1 mm and 3 mm. The first strategy will be discussed in Sect.~\ref{sec:sourcesize} and the second analysis in Sect.~\ref{sec:1mm-3mm}.

\subsubsection{Emitting area} \label{sec:sourcesize}

\begin{figure}
\centering
\includegraphics[trim={0cm 13.7cm 0cm 0.7cm},clip]{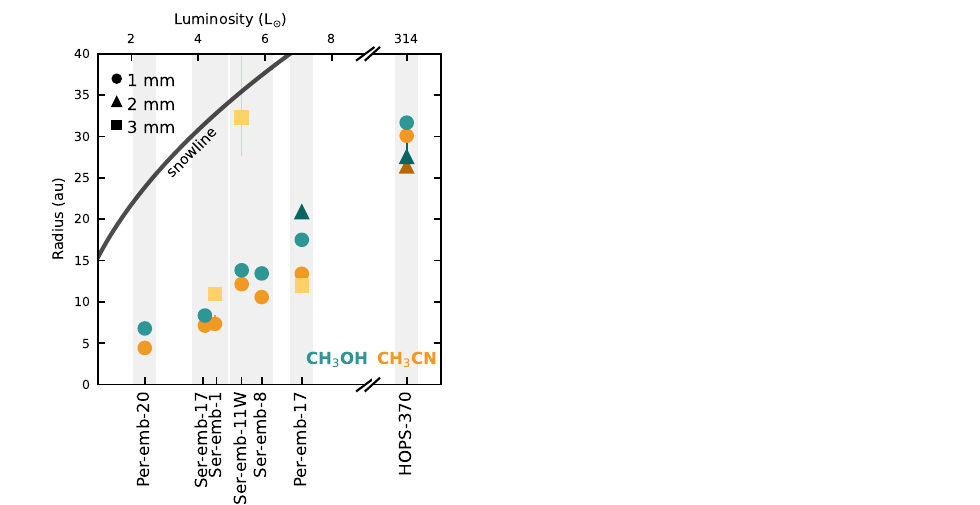}
\caption{Overview of emitting areas derived from CH$_3$CN (orange) and CH$_3$OH (teal) as function of bolometric luminosity. Results from 1 mm observations ($J_K=14_K-13_K$ for CH$_3$CN) are shown as circles, results from 2 mm observations ($J_K=12_K-12_K$ for CH$_3$CN) as triangles, and results from 3 mm observations ($J_K=5_K-4_K$ for CH$_3$CN) as squares. The black line marks the expected water snowline location (100 K) in a prototstellar envelope based on the bolometric luminosity (Eq.~{\ref{eq:snowline}}), as a proxy for the CH$_3$OH and CH$_3$CN snowlines. The vertical gray areas are meant to guide the eye.}
\label{fig:sourcesize}
\end{figure}

The emitting areas derived from the population diagram analysis are listed in Tables~\ref{tab:PD_CH3CN} and \ref{tab:PD_CH3OH}, and shown in Fig.~\ref{fig:sourcesize}. For CH$_3$CN, the emitting areas have radii between $\sim$4--13 au for the low-mass sources, although a larger radius of 32 au is found for Ser-emb-11W at 3 mm, and measures $\sim$30 au for HOPS-370. The emitting area for CH$_3$OH is a few au larger for all sources. The emitting area generally increases for higher luminosities.

If the distribution of CH$_3$OH and CH$_3$CN is governed by thermal desorption, their emission should originate from within the water snowline since their binding energies are very similar to that of water \citep[e.g.,][]{Collings2004,Wakelam2017,Das2018,Ferrero2020,Busch2022,Minissale2022}. We can make an estimate of the snowline location based on the luminosity. \citet{Bisschop2007} has shown that the water snowline radius ($R = 100$ K) in high-mass protostars depend on luminosity via 
\begin{equation}
    R_{\rm{snowline}} = 15.4 \sqrt{L_\star/L_\sun} \enspace \mathrm{au}, \label{eq:snowline}
\end{equation}
and \citet{vantHoff2022} verified that the same relation holds for low-mass sources. 

It is clear from Fig.~\ref{fig:sourcesize} that the expected water snowline location is more extended than the observed CH$_3$CN and CH$_3$OH emission. The snowline is expected at least 20 au further out for the lowest luminosity sources and more than 200 au further out for HOPS-370\footnote{Equation~\ref{eq:snowline} is likely not valid for HOPS-370 given the presence of a $\sim$200 au gas-disk \citep{Tobin2019}, which changes the density and temperature structure compared to that of an envelope. In addition, the freeze-out temperature of water and COMs will be higher ($\sim$ 150 K) at the higher densities in the disk. Nonetheless, the temperature structure derived from detailed radiative transfer modeling by \citet{Tobin2020} still places the snowline outside of $\sim$150 au.}. This could mean that the emission is more compact than expected from thermal desorption. Alternatively, the abundances in the inner region are enhanced over the abundances released from the ice, making the unresolved observations more sensitive to this inner region. The differences in emitting area and excitation temperature then suggest that the enhancement for CH$_3$CN is stronger and/or occuring at smaller radii than for CH$_3$OH.

However, if the continuum emission in the inner region is optically thick, the derived emitting area is likely not a spherical area centered on the source, but rather an annulus with an inner radius equal to the dust $\tau$ = 1 radius (in the extreme case of no emission at all along our line of sight toward the source). The derived emitting area should then thus be considered an effective emitting area, and the radii shown in Fig.~\ref{fig:sourcesize} are smaller than the actual spatial distributions. The much larger radius derived for Ser-emb-11W that is in good agreement with the expected snowline location may then indicate that the dust is optically thin at 3 mm toward this source.

Nevertheless, these results suggest that the distribution of CH$_3$CN is indeed different than that of CH$_3$OH as the dust optical depth is expected to equally affect both molecules, and a larger emitting area for CH$_3$OH due to more emission from smaller radii is inconsistent with the lower excitation temperatures. A similar conclusion is reached by Walls et al. (under review) for HOPS-370 using radiative transfer modeling with a physical structure appropriate for this source \citep{Tobin2020}. Observations with high enough angular resolution to at least resolve the snowline will be able to constrain whether the emission is indeed more extended that suggested by the derived emitting areas and potentially whether CH$_3$OH and CH$_3$CN have different spatial distributions.

\subsubsection{Emission at different wavelengths} \label{sec:1mm-3mm}

Another way to address differences in distribution between CH$_3$CN and CH$_3$OH is by comparing emission at different wavelengths (1 and 3 mm). Due to the lower dust opacity at longer wavelengths, emission close to the star can be visible at these wavelengths while being obscured by the dust at shorter wavelengths (see \citealt{DeSimone2020} for an comparison of NGC1333 IRAS4A at mm and cm wavelengths). For both CH$_3$CN and CH$_3$OH we derive higher column densities at 3 mm than at 1 mm (Fig.~\ref{fig:N_1mm-3mm}), and for Per-emb-17 we can derive rotational temperatures for CH$_3$OH at 1, 2 and 3 mm which increases with wavelength (Fig.~\ref{fig:CH3OH_RD}). Overall, these results thus suggest that we indeed trace gas closer to the protostar at 3 mm. 

\begin{figure}
\centering
\includegraphics[trim={0.4cm 13.4cm 0cm 1cm},clip]{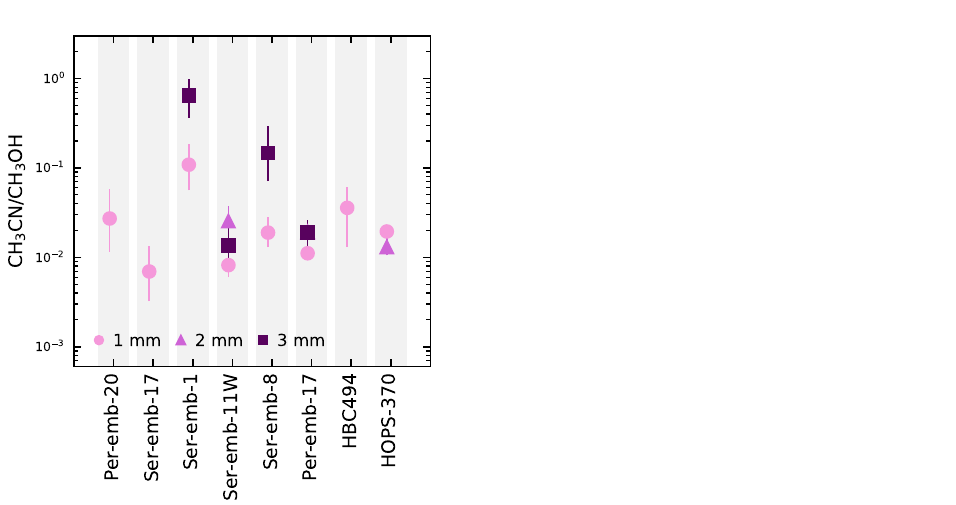}
\caption{CH$_3$CN/CH$_3$OH column density ratio at 1 mm (pink circles), 2 mm (light purple triangles) and 3 mm (dark purple squares) toward the sources in our sample. The error bars take only the fit uncertainty into account since the CH$_3$CN and CH$_3$OH transitions are obtained from the same observations. }
\label{fig:CH3CH-CH3OH}
\end{figure}

However, while the increase in CH$_3$OH column is $\lesssim$~10, the CH$_3$CN column at 3 mm is more than $\sim$10 times higher than at 1 mm toward at least four out of seven sources, with the largest enhancement a factor 46 for Ser-emb-1. This difference is not due to uncertainties in the flux calibration. A flux calibration uncertainty (20\% at 1 and 2 mm, and 10\% at 3 mm) has been included in the uncertainty on the column density, and the column density increase at 3 mm is at least a factor of two. Moreover, the CH$_3$CN and CH$_3$OH column densities at a given wavelength are obtained from the same observations. Nonetheless, to mitigate uncertainties in flux calibration altogether, we present the CH$_3$CN/CH$_3$OH ratio for each source at both wavelengths in Fig.~\ref{fig:CH3CH-CH3OH}. This shows a clear increase in the CH$_3$CN/CH$_3$OH ratio at 3 mm for Ser-emb-1 and Ser-emb-8, and a tentative increase for Per-emb-17 and Ser-emb-11W. In at least four out of the eight sources in our sample (as well as for HOPS-370; Walls et al., subm.) we thus see evidence that CH$_3$CN is enhanced in hot gas close to the protostar compared to CH$_3$OH.

\subsection{Origin of CH$_3$CN enhancement in hot gas} \label{sec:enhancement}

From a chemical perspective, an enhancement of CH$_3$CN in hot gas relative to CH$_3$OH could be due to more effective destruction of CH$_3$CN in warm gas just inside the snowline, more effective destruction of CH$_3$OH in hot gas close to the star, or gas-phase formation of CH$_3$CN in hot gas. While the destruction by H$_3^+$ is approximately three times faster for CH$_3$CN than for CH$_3$OH \citep{McElroy2013}, the short timescales at these high densities ($< 10^5$~yr; \citealt{Hatchell1998}) would mean that a relative change in abundance would be visible only for a short period of time. Given that we observe a difference between CH$_3$CN and CH$_3$OH for both Class 0 and Class I protostars, more efficient gas-phase destruction of CH$_3$CN just inside the snowline is unlikely to be the origin. 

Another destruction mechanism is photodissociation upon the desorption of UV photons, but since photodissociation rates for CH$_3$CN and CH$_3$OH are similar \citep[][and references therein]{Heays2017}, this is not expected to result in an increased CH$_3$CN/CH$_3$OH ratio in hot gas. Because CH$_3$OH formation occurs solely in ices, unlike CH$_3$CN which can form in the gas, the CH$_3$OH abundance is expected to decrease once all ice has sublimated, especially when a disk is present and the radial migration of gas is slow \citep{Aikawa2020}. Although a detailed analysis of the radial abundance would be required for firm conclusions, the 2--10 times higher column densities at 3 mm (Fig.~\ref{fig:N_1mm-3mm}) do not immediately suggest a strong decrease of CH$_3$OH in the inner region. In addition, \citet{Tobin2020} and Walls et al. (subm.) show that the CH$_3$OH emission toward HOPS-370 can be reproduced with a constant abundance profile. More effective gas-phase destruction of CH$_3$OH is thus also unlikely to be the cause of the observed differences. 

The relative enhancement of CH$_3$CN in hot gas is thus probably due to formation of CH$_3$CN at high temperatures. This is, for example, seen in the models by \citet{Garrod2013,Garrod2017,Garrod2022}, where above $\sim$300 K, H atoms can overcome the activation energy barrier to react efficiently with NH$_3$, NH$_2$ and NH, gradually increasing the amount of atomic nitrogen through abstraction of hydrogen. Atomic nitrogen then reacts with CH$_3$ to form HCN and subsequently CH$_3$CN. This process depends on the presence of NH$_3$ (see also, e.g., \citealt{Taquet2016}), whose abundance in these models is substantially higher (17\% with respect to H$_2$O ice) than observed toward low-mass protostars (3--10\%; \citealt{Boogert2015}, and references therein) and dense clouds ($\sim$4-5\%; \citealt{McClure2023}). In addition, a strong increase in CH$_3$CN is only visible in the slow warm-up models, where 300~K is reached after $10^6$ yr, which is longer than the protostellar lifetime of a few $10^5$ yr. For medium warm-up time scales, where 300 K is reached after $2 \times 10^5$ yr, the increase in CH$_3$CN abundance is minimal even though there is a strong increase in HCN. It is therefore not clear whether these models can explain the observations presented here. 

\begin{figure*}
\centering
\includegraphics[trim={0cm 13.8cm 0cm 1.5cm},clip]{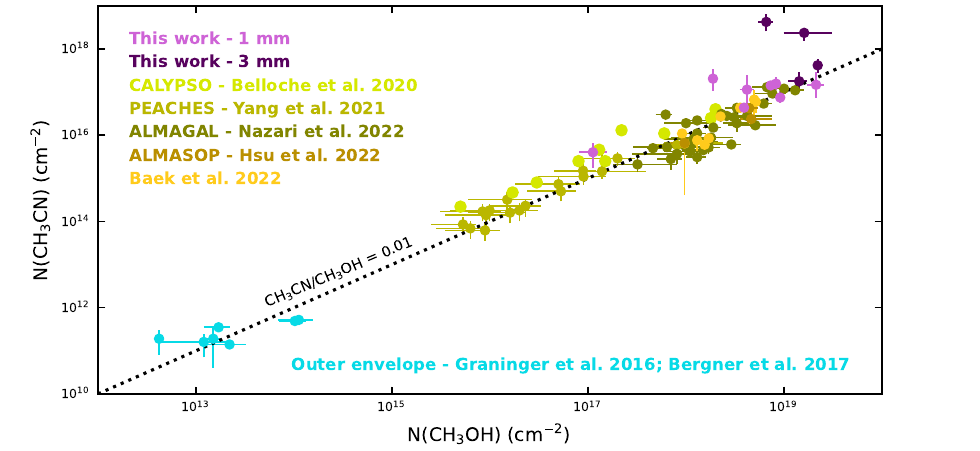}
\caption{CH$_3$CN column densities versus CH$_3$OH column densities as derived in this work at 1 mm (purple) and 3 mm (dark purple) and from surveys of low and high-mass protostars in the recent literature \citep{Graninger2016,Bergner2017,Belloche2020,Yang2021,Nazari2022,Hsu2022,Baek2022}. For reference, the dotted line marks a CH$_3$CN/CH$_3$OH column density ratio of 0.01.}
\label{fig:literature}
\end{figure*}

In the most recent models by \citet{Garrod2022} (labeled as ``final'' in that work), the HCN abundance is two orders of magnitude higher than the CH$_3$CN abundance in $\gtrsim$300 K gas, and both molecules are expected to trace similar regions and conditions. We can therefore test some of these model predictions using the HC$^{15}$N $J=3-2$ transition at 258.156996 GHz covered in our 1 mm data sets (Fig.~\ref{fig:HC15N}). Assuming optically thin emission in LTE and a conservative $^{14}$N/$^{15}$N ratio of 440 as measured for the solar nebula \citep{Marty2011}\footnote{lower $^{14}$N/$^{15}$N ratios of $\sim$80--300 have been observed in the Solar System's rocky planets, comets and meteorites \citep{Mumma2011}, as well as in protoplanetary disks \citep{Guzman2017,Hily-Blant2017}}, the HC$^{15}$N $J=3-2$ peak brightness temperature is expected to be 2.75 times higher than that for CH$_3$CN $14_4-13_4$ at 300 K. In our observations, HC$^{15}$N is fainter than CH$_3$CN $14_4-13_4$ toward HOPS-370, of comparable brightness toward Per-emb-17 and Ser-emb-11W, and only brighter than CH$_3$CN $14_4-13_4$ toward Per-emb-20, Ser-emb-1, Ser-emb-8 and Ser-emb-17. The strongest difference is observed toward Per-emb-20, but HC$^{15}$N is only 1.7 times stronger than CH$_3$CN $14_4-13_4$. The models by Garrod et al. appear thus not directly fully in agreement with our observations. 

Another process that could be responsible for the enhancement of CH$_3$CN in the hot gas, potentially in tandem with the hot-gas chemistry in the Garrod et al. models, is the thermal destruction of carbonaceous grains and the subsequent release of carbon and nitrogen into the hot gas. The chemical models of protoplanetary disks by \citet{Wei2019} show that the additionally available carbon resulting from carbon-grain destruction strongly increases the HCN abundance. Although these authors do not discuss CH$_3$CN, their results suggest that an increase in HCN in hot gas is possible without having all nitrogen initially in NH$_3$. Alternatively, since it is unknown in what form carbon and nitrogen will be released upon grain destruction, the gas-phase NH$_3$ abundance in $\gtrsim$300 K gas may be higher than the ice abundance due to carbon-grain sublimation. A detailed characterization of the chemistry in the warm and hot gas through both observations and models is required to constrain the origin of the CH$_3$CN enhancement.

\subsection{The prevalence of carbon-grain sublimation} \label{sec:othersources}

An enhancement of CH$_3$CN in hot gas was also found by \citet{Nazari2023} for high-mass protostars in the ALMAGAL survey. These authors determined the relative contribution of warm and hot gas by dividing the column density derived from transitions with low and high ($>$ 400 K) upper-level energies. While the low-/high-energy column density ratio for CH$_3$OH was consistent with what would be expected from a uniform abundance within the snowline, the ratio was higher for CH$_3$CN. A similar result was found for the nitrogen-bearing species C$_2$H$_3$CN, while HNCO was similar to CH$_3$OH. A different distribution for CH$_3$CN and CH$_3$OH, with CH$_3$CN enhanced in hot gas, has now thus been found using two different analysis techniques and toward both low-mass (this work) and high-mass sources \citep{Nazari2023}, suggesting that this may be a common feature of protostellar chemistry. 

Even though CH$_3$OH and CH$_3$CN are the most abundant oxygen- and nitrogen-bearing COM species, hot ($\gtrsim$ 300 K) gas traced in particular by CH$_3$CN has not been widely reported toward low-mass protostars. This raises the question whether our sample is biased or whether these signatures were not detectable with other studies. Since observations of COMs, even with ALMA, typically do not have the spatial resolution to resolve the emission, the focus has generally been on detection rates and a comparison of (relative) column densities amongst sources. For example, one of the first chemical surveys targeting a large number of low-mass protostars, CALYPSO, could only derive rotational temperatures for both CH$_3$OH and CH$_3$CN toward four out of the 12 sources with COM emission, because the 3 mm observations covering CH$_3$CN were typically not sensitive enough to detect $\nu_8 = 1$ transitions while the $\nu = 0$ transitions were often optically thick \citep{Belloche2020}. Ser-emb-8 was part of the CALYPSO sample, and the rotational temperature found for CH$_3$OH (210 $\pm$ 37 K) is consistent with the results presented here (231 $\pm$ 3 K). Our rotational temperature of 358~K for CH$_3$CN then suggests that hot gas may have gone undetected with the CALYPSO observations. 

Moreover, high spectral resolution with ALMA typically comes at the cost of reduced bandwidth, which limits the number of transitions detected per source, such that excitation conditions cannot always be constrained. For example, the PEACHES survey of low-mass protostars in Perseus covered only three low-energy CH$_3$CN transitions ($14_K-13_K$, with $K=0-2$), preventing a temperature measurement \citep{Yang2021}. In addition, as the low $K$-transitions can get optically thick in colder gas, hot gas is thus also very likely to go undetected in the PEACHES survey. 

Recent chemical surveys have found a tight correlation between column densities of CH$_3$OH and CH$_3$CN with CH$_3$CN/CH$_3$OH $\sim$0.01 over seven orders of magnitude in column density and for both the outer envelope and the hot core (Fig.~\ref{fig:literature}; \citealt{Bergner2017,Belloche2020,Yang2021,Nazari2022,Hsu2022}). At 1 mm, seven sources in our sample follow this trend very well, while one (Ser-emb-1) is located slightly above it (see also Fig.~\ref{fig:CH3CH-CH3OH}). Our targets are thus not standing out from those in previous samples. They are located at the higher end of the column density ranges because we constrained the emitting area. Interestingly, three sources that show an enhancement of CH$_3$CN in hot gas (Per-emb-17, Ser-emb-8 and Ser-emb-11W) follow the trend, while only one (Ser-emb-1) is slightly above it. The CH$_3$CN/CH$_3$OH ratio at 1 mm alone is thus not informative about the presence of hot gas or a chemical signature of carbon-grain sublimation. This is reinforced by the three targets from the ALMASOP survey of Orion protostars, which all display rotational temperatures of 210--280 K for CH$_3$OH, while higher temperatures of 400--430 K are found for CH$_3$CN (measured from $^{13}$CH$_3$CN; \citealt{Hsu2022}). Hot gas and potential signatures of carbon-grain sublimation may thus have gone undetected in previous surveys and may be common based on the results presented here. 

The ratios derived here at 3 mm (for Ser-emb-1 and Ser-emb-8) lie clearly above the trend, hinting that a CH$_3$CN/CH$_3$OH ratio of $\sim$0.01 may be reflective of the ice abundances, while the ratio increases in hot gas due to additional gas-phase formation of CH$_3$CN. However, detailed studies of larger source samples are required to confirm this as the CH$_3$CN/CH$_3$OH ratio at 3 mm for Per-emb-17 and Ser-emb-11W are close to 0.01, and the CALYPSO observations analysed by \citet{Belloche2020} use 3 mm observations of CH$_3$CN as well (although CH$_3$OH observations at both 2 and 3 mm). The effects of, for example, the spatial extend and opacity of the continuum on the CH$_3$CN/CH$_3$OH ratio would need to be studied in more detail, as the CALYPSO sources with COM detections have bright continuum emission which may shield the hot gas from our view even at 3~mm.


\section{Conclusions}\label{sec:Conclusion}

We have presented spatially unresolved NOEMA observations at 1, 2 and 3 mm of CH$_3$CN and CH$_3$OH toward nine low-mass protostars and one intermediate-mass protostar in different star-forming regions. Both molecules are detected toward seven of the low-mass protostars as well as toward the intermediate-mass protostar. The main findings from our analysis using population diagrams are as follows: 

\begin{itemize}
    \item Not all outbursting protostars display line-rich spectra, as only a few lines from both COMs and simple species are detected toward HBC494. 

    \item Evidence of hot ($\gtrsim$300~K) gas based on CH$_3$CN emission is found toward all eight sources, suggesting that hot gas may be common around (low-mass) protostars. 
    
    \item Rotational temperatures are consistently higher for CH$_3$CN ($\sim$280--425 K) than for CH$_3$OH ($\sim$135--250 K), suggesting CH$_3$CN traces hotter gas than CH$_3$OH. There are no trends with source luminosity, but the intermediate-mass protostar displays the highest CH$_3$CN rotational temperature and an average CH$_3$OH rotational temperature, suggesting CH$_3$CN may be more strongly affected by the presence of hot gas than CH$_3$OH. 

    \item The emitting area of CH$_3$CN is smaller than that for CH$_3$OH, hinting at a more compact distribution for CH$_3$CN or a CH$_3$CN enhancement close to the protostar. 

    \item Higher column densities are traced at 3 mm compared to 1 mm for both CH$_3$CN and CH$_3$OH. In addition, in the only source with enough detections, the CH$_3$OH rotational temperature increased at longer wavelengths. These results suggest that hotter gas closer to the star is traced at longer wavelengths due to lower dust opacities. The increase in column density is stronger for CH$_3$CN than for CH$_3$OH, implying a stronger increase of CH$_3$CN in hot gas.  
    
\end{itemize}

All together, these results suggests that hot $\gtrsim$300 K gas is common around low and intermediate-mass protostars and that optically thin high-$K$ transitions of CH$_3$CN, preferentially at longer wavelengths, are required to trace this gas. If carbon-grain sublimation would be the cause of the enhancement of CH$_3$CN in this hot gas then the early formation conditions of Earth may not be rare.


\acknowledgments 

The authors thank the referee for suggestions that improved the manuscript. M.L.R.H. would like to thank Ryan Loomis and Brett McGuire for discussing the analysis of CH$_3$CN emission and acknowledges support from the Michigan Society of Fellows.


\bibliography{References}{}
\bibliographystyle{aasjournal}


\restartappendixnumbering

\begin{appendix}

\section{NOEMA observing log} 

Details of the NOEMA observations are presented in Table~\ref{tab:NOEMA}.

\begin{deluxetable*}{llccccccc}
\tablecaption{NOEMA observing log. \label{tab:NOEMA}}
\tablewidth{0pt}
\addtolength{\tabcolsep}{-1pt} 
\tabletypesize{\scriptsize}
\tablehead{
\colhead{NOEMA} \vspace{-0.3cm} & \colhead{Sources} & \colhead{Date} & \colhead{Config.} & \colhead{Bandpass} & \colhead{Phase} & \colhead{Flux}  & \colhead{Spectral} & \colhead{Time on source}\\
\colhead{program} & \colhead{} & \colhead{} & \colhead{} & \colhead{calibrator} & \colhead{calibrator} & \colhead{calibrator}  & \colhead{setting$^{a}$} & \colhead{(hour)}
} 
\startdata 
W19AA & HBC494                           & 14 Feb 2020 & 10A & 3C84     & J0542-0913 & LKHA101 & 1 mm & 2.2 \\
      & HCB494                           & 22 Feb 2020 & 10A & 3C84     & J0542-0913 & LKHA101 & 1 mm & 1.1 \\
      & Per-emb-17, Per-emb-20           & 05 Feb 2020 & 10A & 3C84     & 0333+321   & LKHA101 & 1 mm & 0.6, 1.2 \\
      & Per-emb-17, Per-emb-20           & 06 Feb 2020 & 10A & 3C454.3  & 0333+321   & MWC349  & 1 mm & 2.2, 4.5 \\
      & Ser-emb-1, Ser-emb-8, Ser-emb-17 & 07 Feb 2020 & 10A & 3C345    & 1827+062   & MWC349  & 1 mm & 1.0, 1.0, 1.0 \\
      & Ser-emb-1, Ser-emb-8, Ser-emb-17 & 15 Feb 2020 & 10A & 3C273    & 1827+062   & MWC349  & 1 mm & 1.0, 1.0, 1.0 \\
      & Ser-emb-1, Ser-emb-8, Ser-emb-17 & 18 Feb 2020 & 10A & 1749+096 & 1827+062   & MWC349  & 1 mm & 0.8, 0.8, 0.8 \\
      & Ser-emb-1, Ser-emb-8, Ser-emb-17 & 20 Feb 2020 & 10A & 3C84     & 1827+062   & MWC349  & 1 mm & 1.2, 1.2, 1.2 \\
W20AD & HBC494                           & 15 Feb 2021 & 11A & 3C84     & J0542-0913 & LKHA101 & 3 mm & 3.4 \\
      & Per-emb-17, Per-emb-20           & 16 Feb 2021 & 10A & 3C84     & 0333+321, 0322+222  & MWC349 & 3 mm & 2.1, 2.1 \\
      & Ser-emb-1, Ser-emb-8, Ser-emb-17 & 09 Feb 2021 & 11A & 3C345    & 1801+010, 1851+0035 & MWC349 & 3 mm & 0.7, 0.6, 0.5 \\
      & Ser-emb-1, Ser-emb-8, Ser-emb-17 & 17 Feb 2021 & 11A & 3C345    & 1801+010, 1851+0035 & MWC349 & 3 mm & 1.0, 1.0, 1.0 \\
      & Ser-emb-1, Ser-emb-8, Ser-emb-17 & 23 Feb 2021 & 11A & 3C345    & 1801+010, 1851+0035 & 3C345  & 3 mm & 1.2, 1.2, 1.2 \\
W20AE & Per-emb-17, Per-emb-20           & 03 Mar 2021 & 11A & 3C84     & 0333+321, 0322+222  & MWC349 & 2 mm & 0.8, 0.8 \\
W20AF & HOPS-370                         & 23 Dec 2020 & 11C & 3C84     & J0542-0913, 0458-020 & LKHA101 & 1 mm & 3.0 \\
      & HOPS-370                         & 06 Jan 2021 & 11C & 3C84     & J0542-0913, 0458-020 & MWC349  & 1 mm & 1.9 \\
      & HOPS-370                         & 17 Feb 2021 & 10A & 3C84     & J0542-0913, 0458-020 & LKHA101 & 1 mm & 2.6 \\
      & HOPS-370                         & 18 Mar 2021 & 11A & 3C84     & J0542-0913, 0458-020 & LKHA101 & 2 mm & 2.2 \\
S21AH & Ser-emb-1, Ser-emb-8, Ser-emb-17 & 19 Nov 2021 & 9C  & 2013+370 & 1801+010, 1851+0035  & MWC349  & 2 mm & 0.9, 0.9, 1.0 \\
      & Ser-emb-1, Ser-emb-8, Ser-emb-17 & 20 Nov 2021 & 9C  & 3C273    & 1801+010, 1851+0035  & MWC349  & 2 mm & 1.1, 1.1, 1.1 \\
\enddata
\vspace{0.1cm}
\textbf{Notes.}
Ser-emb-11W is observed in the same field of view as Ser-emb-17. 
\vspace{-0.2cm}
\tablenotetext{a}{Each spectral setup consists of two $\sim$8 GHz wide sidebands with 2 MHz spectral resolution, and a set of narrow high spectral resolution windows (62.5 kHz resolution). The 1 mm setting covers the frequency ranges of 240.9--248.6 GHz and 256.4--264.1 GHz, the 2 mm setting covers the frequency ranges of 202.7--210.8 GHz and 218.2--226.3 GHz, and the 3 mm setting covers the frequency ranges of 85.2--93.3 GHz and 100.7--108.8 GHz. Here we focus on the 2 MHz resolution data, except for CH$_3$CN $J=5-4$, where we use a high spectral resolution window between 91.92--92.01 GHz, binned to 2 km s$^{-1}$.}
\end{deluxetable*}


\section{Line fitting} \label{ap:LineFitting}

Individual line fluxes are measured by fitting a Gaussian to the line profile using the CLASS package of GILDAS. For CH$_3$CN, the relative positions of the different $K$-components for each $J$-transition are kept fixed, and one line width is enforced for all $K$-components. The line width fit for the different $J$-transitions is generally consistent per source, especially between $14_K-13_K$ and $12_K-11_K$. The broadest lines are observed toward HOPS-370 and Per-emb-17 ($\sim$11 km s$^{-1}$), and the Serpens sources display the most narrow lines ($\sim$5--6 km s$^{-1}$). The linewidth toward HBC494 and Per-emb-20 is $\sim$9 km s$^{-1}$. The linewidths are typically slightly smaller at 3 mm, probably due to the lower signal-to-noise of these observations. The obtained linewidths and integrated fluxes are listed in Tables~\ref{tab:CH3CNdv} and \ref{tab:CH3CN_parameters}, respectively. 

\begin{deluxetable}{lccc}
\tablecaption{Line widths of the CH$_3$CN emission. \label{tab:CH3CNdv}}
\tablewidth{0pt}
\addtolength{\tabcolsep}{-1pt} 
\tabletypesize{\scriptsize}
\tablehead{
\colhead{Source} \vspace{-0.35cm} & \colhead{1 mm} & \colhead{2mm} & \colhead{3 mm} \\
\colhead{} \vspace{-0.5cm} & \colhead{($J=14-13$)} & \colhead{($J=12-11$)} & \colhead{($J=5-4$)} \\
}
\startdata 
HBC494      & 9.3 $\pm$ 0.9  & \nodata        & $-$   \\
HOPS-370    & 11.5 $\pm$ 0.1 & 11.9 $\pm$ 0.1 & \nodata  \\
Per-emb-17  & 11.1 $\pm$ 0.1 & 10.6 $\pm$ 0.2 & 8.4 $\pm$ 0.4 \\
Per-emb-20  & 9.4 $\pm$ 0.3  & $-$            & $-$ \\
Ser-emb-1   & 5.0 $\pm$ 0.2  & 4.4 $\pm$ 0.4  & 6.8 $\pm$ 0.5  \\
Ser-emb-8   & 6.4 $\pm$ 0.2  & 7.2 $\pm$ 0.8  & 3.6 $\pm$ 0.3 \\
Ser-emb-11W & 5.4 $\pm$ 0.1  & 5.4 $\pm$ 0.3  & 4.0 $\pm$ 0.1  \\
Ser-emb-17  & 6.2 $\pm$ 0.3  & $-$            & $-$   \\
\enddata
\vspace{0.1cm}
\textbf{Notes.} Line widths in km s$^{-1}$ obtained from Gaussian fits. Three dots (\nodata) indicate that a frequency range has not been observed, and a dash ($-$) indicates that no transitions are detected.
\end{deluxetable}

The CH$_3$OH transitions are more spread out in frequency space than individual CH$_3$CN $J$-ladders. Hence, we keep the line center as free parameter when fitting Gaussians to obtain the integrated flux, but fix the line width to the width derived for CH$_3$CN in the same dataset. This aids the fit of weak lines and lines in crowded regions of the spectrum, but does not change the overall results compared to when the linewidth is kept as free parameter. Blended transitions are excluded from the analysis. While we use the high spectral resolution 3 mm spectra for CH$_3$CN $5_K-4_K$, the majority of CH$_3$OH lines in the 3 mm dataset are not covered at high spectral resolution so we use the low-resolution (2 MHz) dataset. For Ser-emb-8 and Ser-emb-11W, the narrow lines are then detected only in 1--2 channels at 3 mm. Instead of a Gaussian fit, we estimate their line flux by multiplying the peak intensity by the CH$_3$CN line width. Integrated fluxes for the observed CH$_3$OH transitions that are not blended and used in the analysis are listed in Table~\ref{tab:CH3OH_parameters}.

Molecular line parameters and the partition function for CH$_3$CN are obtained from the Cologne Database for Molecular Spectroscopy (CDMS; \citealt{Muller2015}), which uses data from \citet{Cazzoli2006} and \citet{Muller2009}. The 2016 version of the CDMS line list for CH$_3$CN treats the double statistical weight of the $K=3,6,9,...$ transitions by listing two entries for these transitions, both with $g_{\rm{up}}$ equal to $g_{\rm{up}}$ of the other $K$-transitions. To properly account for optical depth, we treat the $K=3,6,9,...$ transitions as one transition with double the statistical weight of the other $K$-transitions. This means we replace the two entries for the $K=3,6,9,...$ transitions by one that has the combined intensity and $g_{\rm{up}}$ of the two individual entries. The molecular line parameters are listed in Table~\ref{tab:CH3CN_parameters}. Molecular line parameters for CH$_3$OH are also obtained from CDMS, which are based on data from \citet{Xu2008}. The partition function includes both symmetry states as well as the $\nu_t = 0-2$ torsional states (see Table~\ref{tab:CH3OH_parameters}). 

\begin{longrotatetable}
\begin{deluxetable*}{c r r r c c c c c c c c c}
\tablecaption{Spectroscopic parameters and integrated intensities for the observed CH$_3$CN transitions.
\label{tab:CH3CN_parameters}}
\tablewidth{700pt}
\tabletypesize{\scriptsize}
\tablehead{
\colhead{Transition} \vspace{-0.3cm}& \colhead{Frequency} & \colhead{$E_{\rm{up}}$} & \colhead{$g_{\rm{up}}$} & \colhead{$A_{\rm{ul}}$} & \multicolumn{8}{c}{W} \\ 
\colhead{} \vspace{-0.3cm}& \colhead{(GHz)} & \colhead{(K)} & \colhead{} & \colhead{(log$_{10}$(s$^{-1}$))} & \multicolumn{8}{c}{(K km s$^{-1}$)} \\ 
\colhead{} \vspace{-0.5cm}& \colhead{} & \colhead{} & \colhead{} & \colhead{} & \colhead{HBC494} & \colhead{HOPS-370} & \colhead{Per-emb-17} & \colhead{Per-emb-20} &\colhead{Ser-emb-1} & \colhead{Ser-emb-8} & \colhead{Ser-emb-11W} & \colhead{Ser-emb-17} \\ 
} 
\startdata 
$5_0-4_0$ & 91.987088 & 13 & 22 & -4.20 & $-$ & \nodata & 10.7 $\pm$ 1.2 & $-$ & 3.8 $\pm$ 0.5 & 1.7 $\pm$ 0.4 & 5.1 $\pm$ 0.5 & $-$ \\ 
$5_1-4_1$ & 91.985314 & 20 & 22 & -4.22 & $-$ & \nodata & 5.6 $\pm$ 1.1 & $-$ & 2.1 $\pm$ 0.5 & $-$ & 6.2 $\pm$ 0.5 & $-$ \\ 
$5_2-4_2$ & 91.979995 & 42 & 22 & -4.27 & $-$ & \nodata & 6.0 $\pm$ 1.0 & $-$ & 4.0 $\pm$ 0.6 & 2.6 $\pm$ 0.5 & 6.4 $\pm$ 0.5 & $-$ \\ 
$5_3-4_3$ & 91.971131 & 78 & 44 & -4.39 & $-$ & \nodata & 7.3 $\pm$ 1.1 & $-$ & 3.1 $\pm$ 0.5 & 2.1 $\pm$ 0.5 & 7.1 $\pm$ 0.5 & $-$ \\ 
$5_4-4_4$ & 91.958726 & 128 & 22 & -4.64 & $-$ & \nodata & 3.4 $\pm$ 1.0 & $-$ & 2.2 $\pm$ 0.5 & $-$ & 1.9 $\pm$ 0.5 & $-$ \\ 
$12_0-11_0$ & 220.747262 & 69 & 50 & -3.03 & $-$ & 17.6 $\pm$ 0.8 & 27.0 $\pm$ 2.0 & $-$ & 2.3 $\pm$ 0.7 & 3.1 $\pm$ 1.0 & 4.7 $\pm$ 0.9 & $-$ \\ 
$12_1-11_1$ & 220.743011 & 76 & 50 & -3.04 & $-$ & 24.3 $\pm$ 0.9 & 25.7 $\pm$ 2.1 & $-$ & 5.4 $\pm$ 0.8 & 4.3 $\pm$ 1.2 & 5.3 $\pm$ 1.0 & $-$ \\ 
$12_2-11_2$ & 220.730261 & 97 & 50 & -3.05 & $-$ & 22.9 $\pm$ 0.6 & 28.3 $\pm$ 1.6 & $-$ & 2.7 $\pm$ 0.8 & 5.6 $\pm$ 1.0 & 4.8 $\pm$ 1.0 & $-$ \\ 
$12_3-11_3$ & 220.709017 & 133 & 100 & -3.06 & $-$ & 25.5 $\pm$ 0.6 & 27.4 $\pm$ 1.6 & $-$ & 4.4 $\pm$ 0.8 & 3.5 $\pm$ 1.0 & 5.9 $\pm$ 0.9 & $-$ \\ 
$12_4-11_4$ & 220.679287 & 183 & 50 & -3.09 & $-$ & 17.7 $\pm$ 0.6 & 22.9 $\pm$ 1.6 & $-$ & $-$ & 4.7 $\pm$ 1.0 & 4.0 $\pm$ 0.9 & $-$ \\ 
$12_5-11_5$ & 220.641084 & 247 & 50 & -3.12 & $-$ & 18.6 $\pm$ 1.5 & 15.8 $\pm$ 1.6 & $-$ & $-$ & 3.7 $\pm$ 1.0 & 3.1 $\pm$ 0.9 & $-$ \\ 
$12_6-11_6$ & 220.594424 & 326 & 100 & -3.16 & $-$ & 18.0 $\pm$ 1.9 & blended & $-$ & 1.8 $\pm$ 0.7 & $-$ & 3.5 $\pm$ 0.9 & $-$ \\ 
$12_7-11_7$ & 220.539324 & 419 & 50 & -3.22 & $-$ & 7.7 $\pm$ 0.2 & 4.1 $\pm$ 1.4 & $-$ & $-$ & $-$ & 2.3 $\pm$ 0.8 & $-$ \\ 
$12_8-11_8$ & 220.475808 & 526 & 50 & -3.29 & $-$ & 5.2 $\pm$ 0.5 & $-$ & $-$ & $-$ & $-$ & $-$ & $-$ \\ 
$12_9-11_9$ & 220.403901 & 647 & 100 & -3.39 & $-$ & blended & $-$ & $-$ & $-$ & $-$ & $-$ & $-$ \\ 
$12_{10}-11_{10}$ & 220.323631 & 782 & 50 & -3.55 & $-$ & 2.3 $\pm$ 0.4 & $-$ & $-$ & $-$ & $-$ & $-$ & $-$ \\ 
$14_0-13_0$ & 257.527384 & 93 & 58 & -2.83 & 3.6 $\pm$ 1.8 & 23.0 $\pm$ 0.3 & 65.3 $\pm$ 2.8 & 8.3 $\pm$ 1.2 & 10.9 $\pm$ 0.8 & 18.1 $\pm$ 1.4 & 14.1 $\pm$ 0.8 & 12.6 $\pm$ 1.1 \\ 
$14_1-13_1$ & 257.522428 & 100 & 58 & -2.83 & 4.7 $\pm$ 1.9 & 30.2 $\pm$ 0.2 & 82.9 $\pm$ 2.6 & 11.3 $\pm$ 1.2 & 12.3 $\pm$ 0.9 & 22.7 $\pm$ 1.4 & 17.6 $\pm$ 0.9 & 8.2 $\pm$ 1.1 \\ 
$14_2-13_2$ & 257.507562 & 121 & 58 & -2.84 & 3.5 $\pm$ 1.5 & 28.8 $\pm$ 0.3 & 72.5 $\pm$ 2.1 & 10.0 $\pm$ 1.0 & 8.1 $\pm$ 0.8 & 18.4 $\pm$ 1.4 & 16.8 $\pm$ 0.9 & 11.0 $\pm$ 1.1 \\ 
$14_3-13_3$ & 257.482792 & 157 & 116 & -2.85 & 5.9 $\pm$ 1.7 & 31.9 $\pm$ 0.3 & 82.4 $\pm$ 2.2 & 12.2 $\pm$ 1.0 & 9.3 $\pm$ 0.9 & 21.5 $\pm$ 1.4 & 16.6 $\pm$ 0.8 & 10.0 $\pm$ 1.1 \\ 
$14_4-13_4$ & 257.448128 & 207 & 58 & -2.87 & 2.5 $\pm$ 1.4 & 21.2 $\pm$ 0.3 & 48.8 $\pm$ 2.1 & 5.8 $\pm$ 1.0 & 6.2 $\pm$ 0.8 & 15.3 $\pm$ 1.3 & 9.4 $\pm$ 0.8 & 3.6 $\pm$ 1.0 \\ 
$14_5-13_5$ & 257.403585 & 271 & 58 & -2.89 & $-$ & blended & blended & blended & blended & blended & blended & blended \\ 
$14_6-13_6$ & 257.349180 & 350 & 116 & -2.92 & $-$ & 21.1 $\pm$ 1.2 & 38.9 $\pm$ 2.1 & 7.7 $\pm$ 1.0 & 5.1 $\pm$ 0.9 & 12.0 $\pm$ 1.3 & 12.7 $\pm$ 0.8 & 5.0 $\pm$ 0.6 \\ 
$14_7-13_7$ & 257.284935 & 442 & 58 & -2.96 & $-$ & 9.6 $\pm$ 1.2 & 17.7 $\pm$ 2.0 & $-$ & blended & 6.7 $\pm$ 0.6 & 4.6 $\pm$ 1.0 & 2.0 $\pm$ 0.6 \\ 
$14_8-13_8$ & 257.210878 & 549 & 58 & -3.00 & $-$ & 7.3 $\pm$ 1.2 & 7.9 $\pm$ 2.6 & $-$ & 3.5 $\pm$ 1.1 & 6.5 $\pm$ 1.2 & $-$ & 3.4 $\pm$ 0.8 \\ 
$14_9-13_9$ & 257.127036 & 670 & 116 & -3.06 & $-$ & 9.6 $\pm$ 0.8 & 21.0 $\pm$ 2.7 & $-$ & 4.1 $\pm$ 1.1 & 10.0 $\pm$ 1.2 & 5.1 $\pm$ 0.9 & 4.6 $\pm$ 0.9 \\ 
$14_{10}-13_{10}$ & 257.033444 & 806 & 58 & -3.14 & $-$ & 4.7 $\pm$ 1.1 & $-$ & $-$ & $-$ & $-$ & $-$ & $-$ \\ 
$14_{11}-13_{11}$ & 256.930140 & 955 & 58 & -3.25 & $-$ & 1.6 $\pm$ 0.3 & $-$ & $-$ & $-$ & $-$ & $-$ & $-$ \\ 
\enddata
\vspace{0.1cm}
\textbf{Notes.} Three dots (\nodata) indicate that a transition has not been observed, and a dash ($-$) indicates that a transition has not been detected.
\end{deluxetable*}
\end{longrotatetable}

\begin{longrotatetable}
\begin{deluxetable*}{c r r r c c c c c c c c c}
\tablecaption{Spectroscopic parameters and integrated intensities for CH$_3$OH transitions used in the analysis.
\label{tab:CH3OH_parameters}}
\tablewidth{700pt}
\tabletypesize{\scriptsize}
\tablehead{
\colhead{Transition} \vspace{-0.3cm}& \colhead{Frequency} & \colhead{$E_{\rm{up}}$} & \colhead{$g_{\rm{up}}$} & \colhead{$A_{\rm{ul}}$} & \multicolumn{8}{c}{W} \\ 
\colhead{} \vspace{-0.3cm}& \colhead{(GHz)} & \colhead{(K)} & \colhead{} & \colhead{(log$_{10}$(s$^{-1}$))} & \multicolumn{8}{c}{(K km s$^{-1}$)} \\ 
\colhead{} \vspace{-0.5cm}& \colhead{} & \colhead{} & \colhead{} & \colhead{} & \colhead{HBC494} & \colhead{HOPS-370} & \colhead{Per-emb-17} & \colhead{Per-emb-20} &\colhead{Ser-emb-1} & \colhead{Ser-emb-8} & \colhead{Ser-emb-11W} & \colhead{Ser-emb-17} \\ 
} 
\startdata 
6$_{-2}$ -- 7$_{-1}$, v$_t$=0; E & 85.568131 & 75 & 52 & -5.95 & $-$ & \nodata & 5.6 $\pm$ 1.1 & $-$ & $-$ & 0.9 $\pm$ 0.3 & 1.1 $\pm$ 0.3 & $-$ \\ 
7$_{2}^-$ -- 6$_{3}^-$, v$_t$=0; A & 86.615574 & 103 & 60 & -6.16 & $-$ & \nodata & 4.7 $\pm$ 1.0 & $-$ & $-$ & $-$ & 1.7 $\pm$ 0.3 & $-$ \\ 
7$_{2}^+$ -- 6$_{3}^+$, v$_t$=0; A & 86.902916 & 103 & 60 & -6.16 & $-$ & \nodata & 5.0 $\pm$ 0.7 & $-$ & $-$ & $-$ & 1.8 $\pm$ 0.3 & $-$ \\ 
15$_{3}^+$ -- 14$_{4}^+$, v$_t$=0; A & 88.594787 & 328 & 124 & -5.96 & $-$ & \nodata & 3.1 $\pm$ 1.0 & $-$ & $-$ & $-$ & 1.8 $\pm$ 0.3 & $-$ \\ 
15$_{3}^-$ -- 14$_{4}^-$, v$_t$=0; A & 88.939971 & 328 & 124 & -5.95 & $-$ & \nodata & 6.1 $\pm$ 0.9 & $-$ & $-$ & $-$ & $-$ & $-$ \\ 
8$_{-4}$ -- 9$_{-3}$, v$_t$=0; E & 89.505853 & 172 & 68 & -6.12 & $-$ & \nodata & 5.6 $\pm$ 1.0 & $-$ & $-$ & $-$ & $-$ & $-$ \\ 
1$_{0}$ -- 2$_{1}$, v$_t$=1; E & 93.196672 & 303 & 12 & -5.38 & $-$ & \nodata & $-$ & $-$ & $-$ & $-$ & 2.0 $\pm$ 0.3 & $-$ \\ 
10$_{-2}$ -- 10$_{1}$, v$_t$=0; E & 102.122776 & 154 & 84 & -6.76 & $-$ & \nodata & $-$ & $-$ & $-$ & $-$ & 1.3 $\pm$ 0.3 & $-$ \\ 
11$_{-2}$ -- 11$_{1}$, v$_t$=0; E & 102.658160 & 179 & 92 & -6.58 & $-$ & \nodata & $-$ & $-$ & $-$ & 0.8 $\pm$ 0.3 & 1.4 $\pm$ 0.3 & $-$ \\ 
12$_{-2}$ -- 12$_{1}$, v$_t$=0; E & 103.381258 & 207 & 100 & -6.40 & $-$ & \nodata & 4.3 $\pm$ 0.9 & $-$ & $-$ & $-$ & 1.8 $\pm$ 0.3 & $-$ \\ 
13$_{-3}$ -- 12$_{-4}$, v$_t$=0; E & 104.060634 & 274 & 108 & -5.80 & $-$ & \nodata & 6.6 $\pm$ 1.1 & $-$ & $-$ & 0.7 $\pm$ 0.3 & 4.2 $\pm$ 0.3 & $-$ \\ 
11$_{-1}$ -- 10$_{-2}$, v$_t$=0; E & 104.300337 & 159 & 92 & -5.71 & $-$ & \nodata & 12.9 $\pm$ 1.0 & $-$ & $-$ & 0.7 $\pm$ 0.3 & 5.0 $\pm$ 0.3 & $-$ \\ 
13$_{-2}$ -- 13$_{1}$, v$_t$=0; E & 104.336667 & 237 & 108 & -6.23 & $-$ & \nodata & 7.3 $\pm$ 0.9 & $-$ & $-$ & 0.7 $\pm$ 0.3 & 2.9 $\pm$ 0.3 & $-$ \\ 
10$_{4}^-$ -- 11$_{3}^-$, v$_t$=0; A & 104.354831 & 208 & 84 & -5.81 & $-$ & \nodata & 8.3 $\pm$ 0.9 & $-$ & $-$ & 0.9 $\pm$ 0.3 & 2.3 $\pm$ 0.3 & $-$ \\ 
10$_{4}^+$ -- 11$_{3}^+$, v$_t$=0; A & 104.410457 & 208 & 84 & -5.80 & $-$ & \nodata & 8.7 $\pm$ 0.9 & $-$ & $-$ & 0.9 $\pm$ 0.3 & 4.2 $\pm$ 0.3 & $-$ \\ 
13$_{1}^+$ -- 12$_{2}^+$, v$_t$=0; A & 105.063816 & 224 & 108 & -5.67 & $-$ & \nodata & 10.0 $\pm$ 1.0 & $-$ & 2.1 $\pm$ 0.5 & 1.3 $\pm$ 0.3 & 3.6 $\pm$ 0.3 & $-$ \\ 
14$_{-2}$ -- 14$_{1}$, v$_t$=0; E & 105.576396 & 270 & 116 & -6.07 & $-$ & \nodata & 4.8 $\pm$ 1.0 & $-$ & $-$ & $-$ & 2.0 $\pm$ 0.3 & $-$ \\ 
3$_{1}^+$ -- 4$_{0}^+$, v$_t$=0; A & 107.013831 & 28 & 28 & -5.21 & $-$ & \nodata & 20.3 $\pm$ 0.9 & $-$ & 2.7 $\pm$ 0.6 & 1.9 $\pm$ 0.3 & 4.0 $\pm$ 0.3 & 3.9 $\pm$ 0.9 \\ 
15$_{-2}$ -- 15$_{1}$, v$_t$=0; E & 107.159906 & 305 & 124 & -5.92 & $-$ & \nodata & 5.6 $\pm$ 1.0 & $-$ & $-$ & $-$ & 2.0 $\pm$ 0.3 & $-$ \\ 
1$_{1}^+$ -- 2$_{0}^+$, v$_t$=0; A & 205.791270 & 17 & 12 & -4.47 & $-$ & 13.7 $\pm$ 0.4 & 22.9 $\pm$ 1.3 & $-$ & $-$ & $-$ & $-$ & 3.6 $\pm$ 0.9 \\ 
12$_{5}$ -- 13$_{4}$, v$_t$=0; E & 206.001302 & 317 & 100 & -4.98 & $-$ & 8.4 $\pm$ 1.1 & 13.7 $\pm$ 1.5 & $-$ & $-$ & $-$ & 3.6 $\pm$ 0.7 & $-$ \\ 
19$_{1}$ -- 19$_{0}$, v$_t$=0; E & 209.518804 & 462 & 156 & -4.50 & $-$ & 13.1 $\pm$ 0.9 & 16.3 $\pm$ 1.4 & $-$ & $-$ & $-$ & 5.3 $\pm$ 0.8 & $-$ \\ 
4$_{2}$ -- 3$_{1}$, v$_t$=0; E & 218.440063 & 46 & 36 & -4.33 & $-$ & 12.8 $\pm$ 0.6 & 47.4 $\pm$ 1.4 & 3.5 $\pm$ 1.2 & $-$ & $-$ & 5.0 $\pm$ 0.7 & blended \\ 
8$_{0}$ -- 7$_{1}$, v$_t$=0; E & 220.078561 & 97 & 68 & -4.60 & $-$ & 16.0 $\pm$ 0.6 & 39.5 $\pm$ 1.6 & 3.7 $\pm$ 1.0 & $-$ & $-$ & 4.0 $\pm$ 0.8 & 3.0 $\pm$ 0.8 \\ 
16$_{2}^+$ -- 15$_{1}^+$, v$_t$=1; A & 222.722856 & 613 & 132 & -4.44 & $-$ & 4.4 $\pm$ 0.6 & 6.6 $\pm$ 1.6 & $-$ & $-$ & $-$ & 2.7 $\pm$ 0.8 & $-$ \\ 
20$_{-2}$ -- 19$_{-3}$, v$_t$=0; E & 224.699408 & 514 & 164 & -4.70 & $-$ & 8.5 $\pm$ 0.5 & 12.3 $\pm$ 1.3 & $-$ & $-$ & $-$ & 4.2 $\pm$ 0.9 & $-$ \\ 
5$_{3}^-$ -- 4$_{3}^-$, v$_t$=2; A & 240.916172 & 693 & 44 & -4.41 & $-$ & 2.9 $\pm$ 0.2 *& 2.9 $\pm$ 0.4 *& $-$ & $-$ & 3.4 $\pm$ 0.3 *& 3.8 $\pm$ 0.4 *& 2.8 $\pm$ 0.5 *\\ 
5$_{3}^+$ -- 4$_{3}^+$, v$_t$=2; A & 240.916173 & 693 & 44 & -4.41 & $-$ & 2.9 $\pm$ 0.2 *& 2.9 $\pm$ 0.4 *& $-$ & $-$ & 3.4 $\pm$ 0.3 *& 3.8 $\pm$ 0.4 *& 2.8 $\pm$ 0.5 *\\ 
5$_{-1}$ -- 4$_{-1}$, v$_t$=1; E & 241.238144 & 448 & 44 & -4.24 & $-$ & 11.5 $\pm$ 0.7 & 17.2 $\pm$ 1.6 & 4.4 $\pm$ 0.8 & $-$ & 9.7 $\pm$ 1.3 & 9.7 $\pm$ 1.0 & $-$ \\ 
5$_{0}^+$ -- 4$_{0}^+$, v$_t$=1; A & 241.267862 & 458 & 44 & -4.22 & $-$ & 10.8 $\pm$ 0.6 & 15.0 $\pm$ 1.5 & $-$ & $-$ & 6.1 $\pm$ 1.2 & 10.3 $\pm$ 0.9 & 6.8 $\pm$ 1.1 \\ 
5$_{1}^-$ -- 4$_{1}^-$, v$_t$=2; A & 241.364143 & 718 & 44 & -4.24 & $-$ & blended & $-$ & $-$ & $-$ & $-$ & 6.7 $\pm$ 0.9 & $-$ \\ 
5$_{1}^-$ -- 4$_{1}^-$, v$_t$=1; A & 241.441270 & 360 & 44 & -4.24 & $-$ & 13.5 $\pm$ 0.6 & 26.6 $\pm$ 1.4 & 3.8 $\pm$ 0.8 & 3.1 $\pm$ 0.8 & 11.7 $\pm$ 1.2 & 14.0 $\pm$ 0.9 & $-$ \\ 
25$_{3}^-$ -- 25$_{2}^+$, v$_t$=0; A & 241.588758 & 804 & 204 & -4.09 & $-$ & 7.4 $\pm$ 0.6 & 11.6 $\pm$ 1.9 & $-$ & $-$ & 5.3 $\pm$ 1.2 & 10.2 $\pm$ 0.9 & $-$ \\ 
5$_{0}$ -- 4$_{0}$, v$_t$=0; E & 241.700159 & 48 & 44 & -4.22 & 4.4 $\pm$ 0.8 & blended & blended & blended & blended & 29.1 $\pm$ 1.8 & blended & blended \\ 
5$_{-1}$ -- 4$_{-1}$, v$_t$=0; E & 241.767234 & 40 & 44 & -4.24 & 5.7 $\pm$ 1.0 & blended & blended & blended & blended & 43.5 $\pm$ 4.9 & blended & blended \\ 
5$_{0}^+$ -- 4$_{0}^+$, v$_t$=0; A & 241.791352 & 35 & 44 & -4.22 & 4.9 $\pm$ 1.0 & 25.4 $\pm$ 1.2 & 115.8 $\pm$ 1.3 & 16.4 $\pm$ 1.1 & 10.7 $\pm$ 1.0 & 43.8 $\pm$ 5.0 & 19.7 $\pm$ 1.0 & 9.0 $\pm$ 1.6 \\ 
5$_{4}^-$ -- 4$_{4}^-$, v$_t$=0; A & 241.806524 & 115 & 44 & -4.66 & $-$ & blended & blended & blended & 2.5 $\pm$ 0.5 *& blended & blended & blended \\ 
5$_{4}^+$ -- 4$_{4}^+$, v$_t$=0; A & 241.806525 & 115 & 44 & -4.66 & $-$ & blended & blended & blended & 2.5 $\pm$ 0.5 *& blended & blended & blended \\ 
5$_{-4}$ -- 4$_{-4}$, v$_t$=0; E & 241.813255 & 123 & 44 & -4.66 & $-$ & blended & blended & blended & 4.3 $\pm$ 1.0 & blended & blended & blended \\ 
5$_{1}$ -- 4$_{1}$, v$_t$=0; E & 241.879025 & 56 & 44 & -4.22 & $-$ & blended & blended & blended & 8.0 $\pm$ 0.9 & blended & blended & blended \\ 
5$_{2}^+$ -- 4$_{2}^+$, v$_t$=0; A & 241.887674 & 72 & 44 & -4.29 & 3.7 $\pm$ 0.5 & blended & blended & blended & 7.0 $\pm$ 0.9 & blended & blended & blended \\ 
14$_{-1}$ -- 13$_{-2}$, v$_t$=0; E & 242.446084 & 249 & 116 & -4.64 & $-$ & 21.3 $\pm$ 1.3 & 51.0 $\pm$ 2.6 & 8.3 $\pm$ 0.9 & 2.8 $\pm$ 0.9 & blended & 15.6 $\pm$ 1.3 & 7.9 $\pm$ 1.1 \\ 
24$_{3}^-$ -- 24$_{2}^+$, v$_t$=0; A & 242.490245 & 746 & 196 & -4.09 & $-$ & 9.3 $\pm$ 1.3 & 13.9 $\pm$ 2.6 & $-$ & $-$ & $-$ & 10.3 $\pm$ 1.2 & $-$ \\ 
18$_{6}^-$ -- 19$_{5}^-$, v$_t$=0; A & 243.397393 & 590 & 148 & -4.70 & $-$ & 4.2 $\pm$ 0.1 *& 7.2 $\pm$ 1.1 *& $-$ & $-$ & 3.7 $\pm$ 0.6 *& 5.2 $\pm$ 0.5 *& $-$ \\ 
18$_{6}^+$ -- 19$_{5}^+$, v$_t$=0; A & 243.397654 & 590 & 148 & -4.70 & $-$ & 4.2 $\pm$ 0.1 *& 7.2 $\pm$ 1.1 *& $-$ & $-$ & 3.7 $\pm$ 0.6 *& 5.2 $\pm$ 0.5 *& $-$ \\ 
23$_{3}^-$ -- 23$_{2}^+$, v$_t$=0; A & 243.412610 & 690 & 188 & -4.09 & $-$ & 9.5 $\pm$ 0.2 & 16.3 $\pm$ 2.2 & $-$ & $-$ & $-$ & 10.6 $\pm$ 1.1 & $-$ \\ 
5$_{1}^-$ -- 4$_{1}^-$, v$_t$=0; A & 243.915788 & 50 & 44 & -4.22 & $-$ & 26.8 $\pm$ 0.7 & 101.0 $\pm$ 1.7 & 13.9 $\pm$ 1.1 & 6.9 $\pm$ 0.9 & 20.9 $\pm$ 1.1 & 19.2 $\pm$ 1.1 & 11.9 $\pm$ 0.8 \\ 
18$_{-6}$ -- 17$_{-7}$, v$_t$=1; E & 245.094503 & 889 & 148 & -4.68 & $-$ & blended & $-$ & $-$ & $-$ & 3.6 $\pm$ 0.7 & $-$ & $-$ \\ 
21$_{3}^-$ -- 21$_{2}^+$, v$_t$=0; A & 245.223019 & 586 & 172 & -4.08 & $-$ & 13.8 $\pm$ 1.0 & 28.9 $\pm$ 1.5 & 4.6 $\pm$ 0.9 & $-$ & 10.5 $\pm$ 0.9 & 15.3 $\pm$ 0.9 & $-$ \\ 
20$_{3}^-$ -- 20$_{2}^+$, v$_t$=0; A & 246.074605 & 537 & 164 & -4.08 & $-$ & 17.7 $\pm$ 1.0 & 36.1 $\pm$ 1.3 & 6.0 $\pm$ 0.9 & $-$ & 13.4 $\pm$ 0.8 & 18.2 $\pm$ 0.8 & 8.2 $\pm$ 0.9 \\ 
19$_{3}^-$ -- 19$_{2}^+$, v$_t$=0; A & 246.873301 & 491 & 156 & -4.08 & $-$ & 17.7 $\pm$ 1.0 & 41.0 $\pm$ 1.9 & 8.9 $\pm$ 0.9 & 4.2 $\pm$ 0.6 & 9.8 $\pm$ 0.8 & 14.0 $\pm$ 0.8 & 7.7 $\pm$ 1.0 \\ 
16$_{2}$ -- 15$_{3}$, v$_t$=0; E & 247.161950 & 338 & 132 & -4.59 & $-$ & 19.2 $\pm$ 1.4 & 42.7 $\pm$ 1.4 & 5.1 $\pm$ 0.8 & $-$ & blended & 14.2 $\pm$ 0.8 & 8.2 $\pm$ 0.9 \\ 
4$_{2}^+$ -- 5$_{1}^+$, v$_t$=0; A & 247.228587 & 61 & 36 & -4.67 & $-$ & 21.9 $\pm$ 1.4 & 64.8 $\pm$ 1.4 & 5.5 $\pm$ 0.8 & $-$ & 15.5 $\pm$ 0.9 & 14.0 $\pm$ 0.8 & 7.9 $\pm$ 0.9 \\ 
18$_{3}^-$ -- 18$_{2}^+$, v$_t$=0; A & 247.610918 & 447 & 148 & -4.08 & $-$ & 19.6 $\pm$ 1.4 & 49.0 $\pm$ 2.2 & 7.2 $\pm$ 0.6 & $-$ & 11.3 $\pm$ 0.9 & 13.4 $\pm$ 0.7 & 7.8 $\pm$ 0.6 \\ 
12$_{-2}$ -- 13$_{-3}$, v$_t$=1; E & 247.840050 & 545 & 100 & -4.20 & $-$ & 12.7 $\pm$ 0.7 & 27.1 $\pm$ 1.7 & $-$ & $-$ & blended & 12.2 $\pm$ 0.9 & $-$ \\ 
23$_{1}$ -- 23$_{0}$, v$_t$=0; E & 247.968119 & 661 & 188 & -4.35 & $-$ & 8.6 $\pm$ 0.9 & 13.8 $\pm$ 1.7 & $-$ & $-$ & 5.6 $\pm$ 0.7 & 8.7 $\pm$ 0.9 & $-$ \\ 
17$_{3}^-$ -- 17$_{2}^+$, v$_t$=0; A & 248.282424 & 405 & 140 & -4.08 & $-$ & 20.3 $\pm$ 0.3 & 59.2 $\pm$ 1.9 & 8.3 $\pm$ 0.8 & $-$ & 12.6 $\pm$ 1.1 & 13.6 $\pm$ 1.0 & 7.5 $\pm$ 0.9 \\ 
19$_{3}^+$ -- 19$_{2}^-$, v$_t$=0; A & 258.780248 & 491 & 156 & -4.05 & $-$ & 13.3 $\pm$ 0.5 & 50.4 $\pm$ 1.9 & 8.1 $\pm$ 1.0 & $-$ & blended & 14.3 $\pm$ 1.0 & blended \\ 
17$_{2}^-$ -- 16$_{1}^-$, v$_t$=1; A & 259.273686 & 653 & 140 & -4.25 & $-$ & 8.3 $\pm$ 0.7 & 18.4 $\pm$ 2.0 & $-$ & $-$ & 7.5 $\pm$ 1.0 & 9.0 $\pm$ 0.9 & $-$ \\ 
24$_{1}$ -- 24$_{0}$, v$_t$=0; E & 259.581398 & 717 & 196 & -4.31 & $-$ & 6.8 $\pm$ 1.4 & 14.6 $\pm$ 1.6 & $-$ & $-$ & blended & blended & $-$ \\ 
20$_{-8}$ -- 21$_{-7}$, v$_t$=0; E & 260.064318 & 808 & 164 & -4.67 & $-$ & $-$ & $-$ & $-$ & $-$ & 7.2 $\pm$ 1.1 & $-$ & $-$ \\ 
20$_{3}^+$ -- 20$_{2}^-$, v$_t$=0; A & 260.381463 & 537 & 164 & -4.04 & $-$ & 15.4 $\pm$ 1.2 & 52.7 $\pm$ 2.3 & 7.7 $\pm$ 1.0 & 5.5 $\pm$ 0.9 & 18.2 $\pm$ 1.4 & 15.6 $\pm$ 1.2 & 13.0 $\pm$ 0.9 \\ 
21$_{-4}$ -- 20$_{-5}$, v$_t$=0; E & 261.061320 & 624 & 172 & -4.52 & $-$ & 7.4 $\pm$ 1.1 & 12.5 $\pm$ 2.0 & $-$ & $-$ & 4.2 $\pm$ 0.8 & 5.7 $\pm$ 0.8 & $-$ \\ 
12$_{6}$ -- 13$_{5}$, v$_t$=0; E & 261.704409 & 360 & 100 & -4.75 & $-$ & blended & blended & $-$ & $-$ & 8.2 $\pm$ 1.1 & 9.6 $\pm$ 1.0 & $-$ \\ 
2$_{1}$ -- 1$_{0}$, v$_t$=0; E & 261.805675 & 28 & 20 & -4.25 & $-$ & 19.0 $\pm$ 0.3 & 99.6 $\pm$ 2.2 & 12.1 $\pm$ 1.0 & blended & 17.7 $\pm$ 0.9 & 14.8 $\pm$ 1.2 & 10.1 $\pm$ 1.3 \\ 
21$_{3}^+$ -- 21$_{2}^-$, v$_t$=0; A & 262.223872 & 586 & 172 & -4.03 & $-$ & 11.8 $\pm$ 0.9 & 40.0 $\pm$ 1.9 & 5.1 $\pm$ 0.7 & $-$ & 12.4 $\pm$ 1.0 & 8.6 $\pm$ 0.9 & 6.4 $\pm$ 0.7 \\ 
\enddata
\vspace{0.1cm}
\textbf{Notes.} Only transitions that are detected and not blended toward at least one source in the sample are listed. Three dots (\nodata) indicate that a transition has not been observed, and a dash ($-$) indicates that a transition has not been detected. Integrated intensities marked with an asterisk (*) are obtained by dividing the integrated intensity for two blended transitions with equal spectroscopic parameters by two. 
\end{deluxetable*}
\end{longrotatetable}

\section{Rotation and population diagrams} \label{ap:RDandPD}

\begin{figure*}
\centering
\includegraphics[width=\linewidth,trim={0.2cm 6.5cm 1cm 1.1cm},clip]{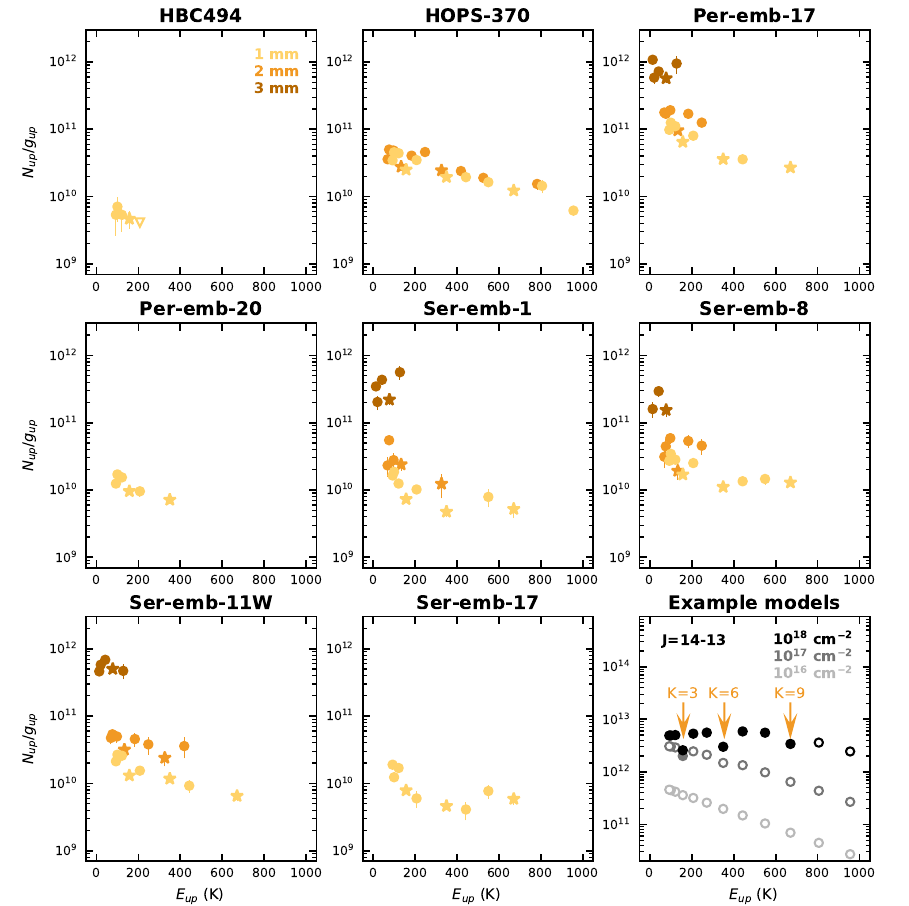}
\caption{Rotation diagrams for the CH$_3$CN $J=5-4$ (dark orange), $J=12-11$ (orange), and $J=14-13$ (yellow) ladders toward the sources in our sample as indicated above each panel. The $K=3$, $K=6$ and $K=9$ transitions are shown as star symbols, the other transitions are shown as circles. The unresolved $12_K-11_K$ and $5_K-4_K$ fluxes have been scaled to the smaller beam size of the unresolved $14_K-13_K$ observations. The vertical scale is the same for all panels, but the beam size differs per source (see Table~\ref{tab:Observations}). The error bars correspond to 1$\sigma$, and are typically smaller than the symbols. The bottom right panel displays three LTE models for the $J=14-13$ ladder at 300 K with varying column densities (different shades of gray), highlighting the behavior of the $K=3$, $K=6$ and $K=9$ transitions when the emission becomes optically thick. Optically thin transitions in this panel are shown as open symbols, and optically thick transitions as filled symbols.} 
\label{fig:CH3CN_RD_total}
\end{figure*}

\begin{figure*}
\centering
\includegraphics[width=\linewidth,trim={0.2cm 6.5cm 1cm 1.1cm},clip]{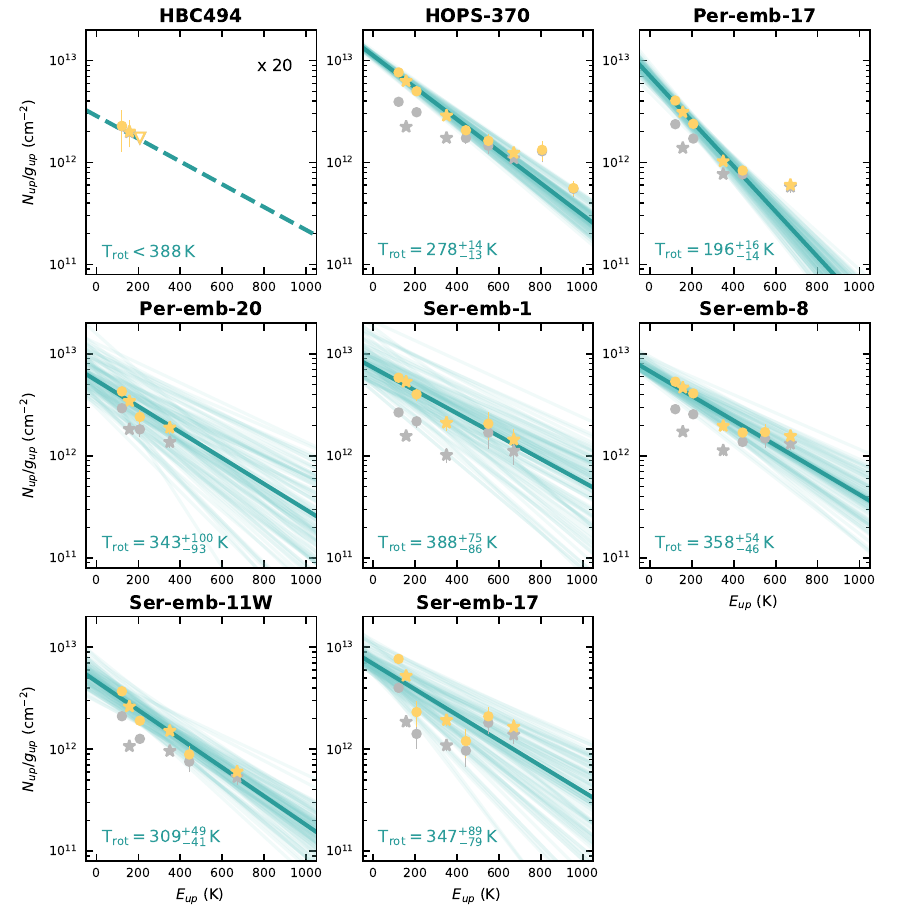}
\caption{Population diagram analysis for CH$_3$CN $J=14-13$. Data points corrected for source size are shown in grey, data points corrected for source size and optical depth are shown in yellow. The blended $K=0$ and $K=1$ transitions are excluded in the modeling and not depicted here. $K=3$, $K=6$ and $K=9$ transitions are shown as star symbols, the other transitions are shown as circles. For HBC494, the values are multiplied by a factor 20 to allow the same vertical scale for all panels, and the triangle denotes the marginal detection of the $K=4$ transition. The error bars correspond to 1$\sigma$, and are typically smaller than the symbols. The rotational temperature, column density and source size were fitted, except for HBC494 were the source size was fixed. Draws from the fit posteriors are shown with light teal lines, with the 50th percentiles shown as thick teal lines. The corresponding rotational temperature is listed in the bottom left corner. All fit results are presented in Table~\ref{tab:PD_CH3CN}.} 
\label{fig:CH3CN_PD_14-13}
\end{figure*}

\begin{figure*}
\centering
\includegraphics[width=\linewidth,trim={0.2cm 17.3cm 1cm 1.1cm},clip]{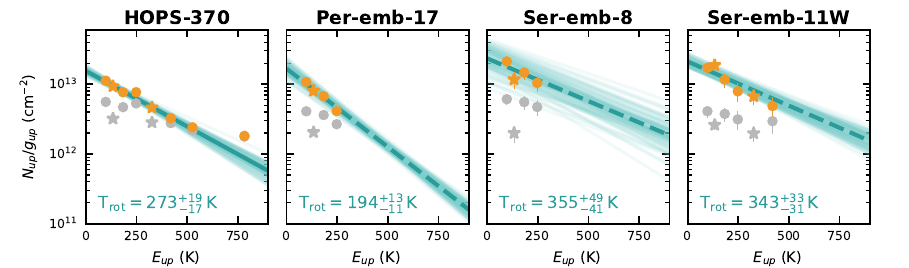}
\caption{As Fig.~\ref{fig:CH3CN_PD_14-13} but for CH$_3$CN $J=12-11$. Solid thick teal lines indicate that the rotational temperature, column density and source size were included in the fit, while a dashed line indicates that the source size was fixed. More details and all fit results are presented in Table~\ref{tab:PD_CH3CN}.} 
\label{fig:CH3CN_PD_12-11}
\end{figure*}

\begin{figure*}
\centering
\includegraphics[width=\linewidth,trim={0.2cm 17.3cm 1cm 1.1cm},clip]{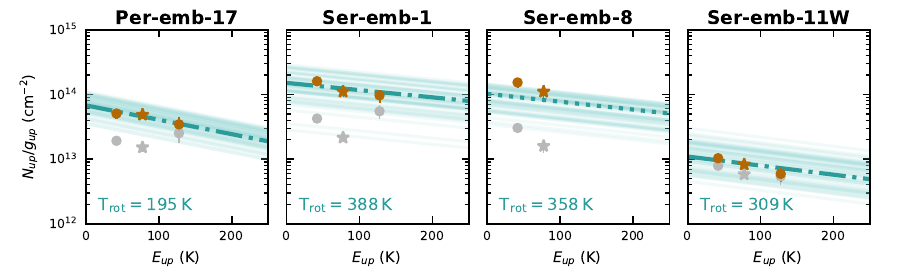}
\caption{As Fig.~\ref{fig:CH3CN_PD_14-13} but for CH$_3$CN $J=5-4$. Dash-dotted lines indicate that the column density and source size were included in the fit while the rotational temperature was fixed, while dotted lines indicate that both the rotational temperature and source size were fixed. More details and all fit results are presented in Table~\ref{tab:PD_CH3CN}.} 
\label{fig:CH3CN_PD_5-4}
\end{figure*}

\begin{figure*}
\centering
\includegraphics[width=\linewidth,trim={0.2cm 6.5cm 1cm 1.1cm},clip]{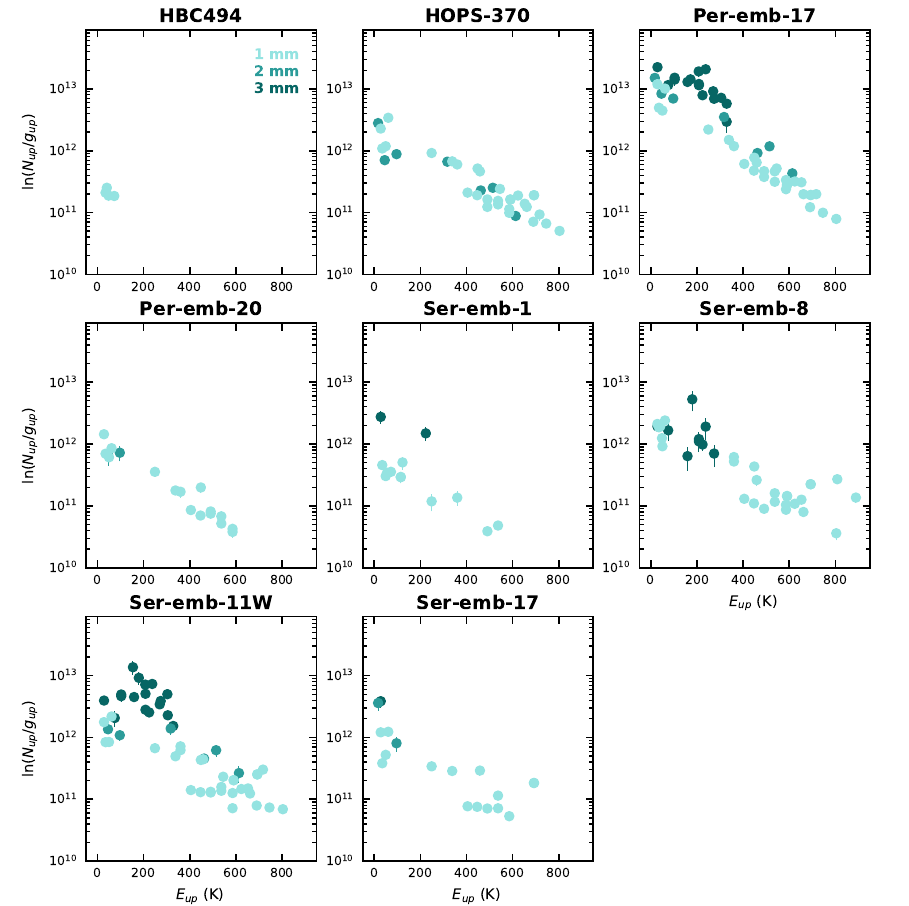}
\caption{Rotation diagrams for CH$_3$OH at 1 mm (light teal), 2 mm (teal) and 3 mm (dark teal) toward the sources in our sample as indicated above each panel. The unresolved fluxes at 2 and 3 mm have been scaled to the smaller beam size of the 1 mm observations. The vertical scale is the same for all panels, but the beam size differs per source (see Table~\ref{tab:Observations}). The error bars correspond to 1$\sigma$, and are typically smaller than the symbols.}
\label{fig:CH3OH_RD_total}
\end{figure*}

\begin{figure*}
\centering
\includegraphics[width=\linewidth,trim={0.2cm 6.5cm 1cm 1.1cm},clip]{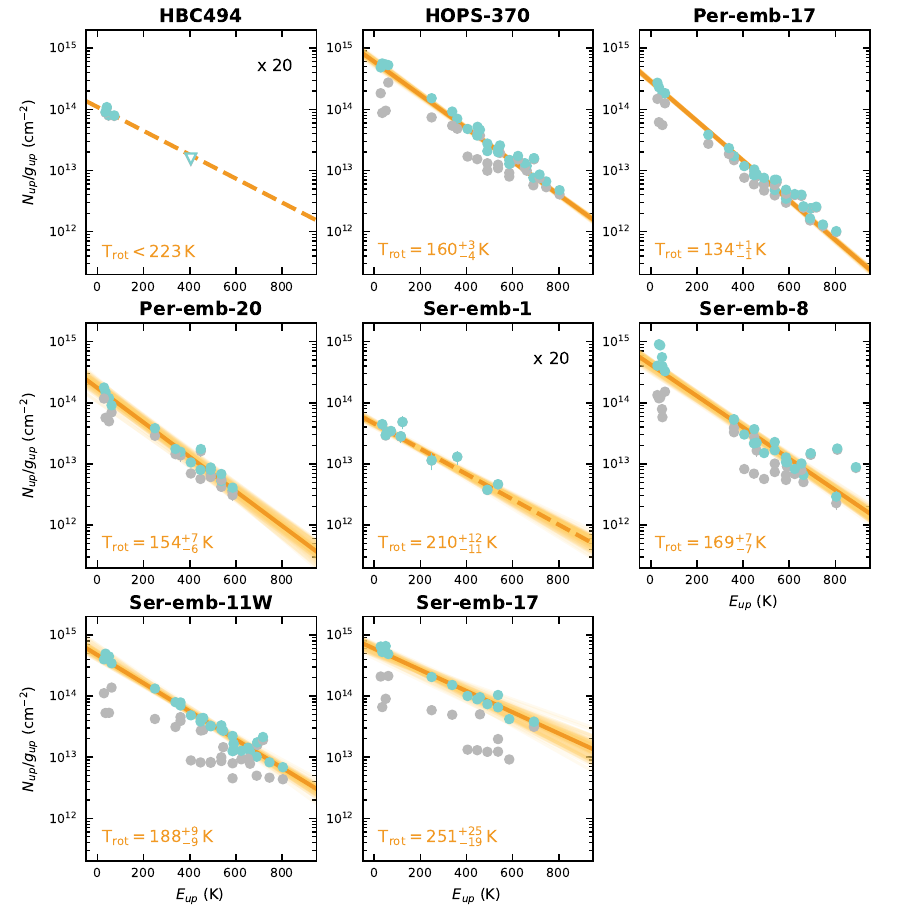}
\caption{As Fig.~\ref{fig:CH3CN_PD_14-13} but for the 1 mm CH$_3$OH transitions. Values for HBC494 and Ser-emb-1 have been multiplied by 20 to allow the same vertical scale for all panels. For HBC494, the triangle denotes the most constraining upper limit of all covered transitions. Solid thick orange lines indicate that the rotational temperature, column density and source size were included in the fit, while a dashed line indicates that the source size was fixed. More details and all fit results are presented in Table~\ref{tab:PD_CH3OH}.} 
\label{fig:CH3OH_PD_1mm}
\end{figure*}

\begin{figure*}
\centering
\includegraphics[width=\linewidth,trim={0.2cm 17.3cm 1cm 1.1cm},clip]{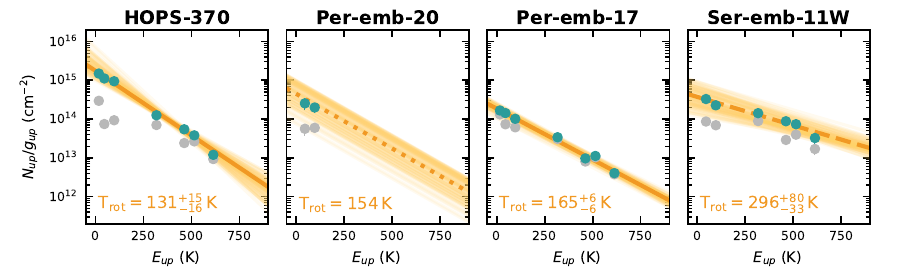}
\caption{As Fig.~\ref{fig:CH3CN_PD_14-13} but for the 2 mm CH$_3$OH transitions. Solid thick orange lines indicate that the rotational temperature, column density and source size were included in the fit, while a dashed line indicates that the source size was fixed and a dotted line means that both the source size and rotational temperature were fixed. More details and all fit results are presented in Table~\ref{tab:PD_CH3OH}.} 
\label{fig:CH3OH_PD_2mm}
\end{figure*}

\begin{figure*}
\centering
\includegraphics[width=\linewidth,trim={0.2cm 17.3cm 1cm 1.1cm},clip]{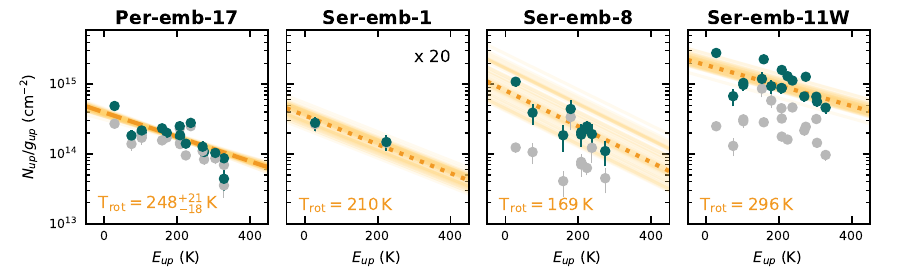}
\caption{As Fig.~\ref{fig:CH3CN_PD_14-13} but for the 3 mm CH$_3$OH transitions. Values for Ser-emb-1 have been multiplied by 20 to allow the same vertical scale for all panels. Dashed lines indicate that the rotational temperature and column density were included in the fit while the source size was fixed. Dotted lines indicate that both the rotational temperature and source size were fixed. More details and all fit results are presented in Table~\ref{tab:PD_CH3OH}.} 
\label{fig:CH3OH_PD_3mm}
\end{figure*}

Rotation diagrams for CH$_3$CN are shown in Fig.~\ref{fig:CH3CN_RD_total}, and population diagrams for the $14_K-13_K$, $14_K-13_K$ and $14_K-13_K$ transitions are presented in Fig.~\ref{fig:CH3CN_PD_14-13}, Fig.~\ref{fig:CH3CN_PD_12-11} and Fig.~\ref{fig:CH3CN_PD_5-4}, respectively. Rotation diagrams for CH$_3$OH are shown in Fig.~\ref{fig:CH3OH_RD_total}, and population diagrams for the emission at 1 mm, 2 mm and 3 mm are displayed in Fig.~\ref{fig:CH3OH_PD_1mm}, Fig.~\ref{fig:CH3OH_PD_2mm} and Fig.~\ref{fig:CH3OH_PD_3mm}, respectively.

\section{Additional figures} 

\begin{figure*}
\subfloat{\includegraphics[trim={0.3cm 14.8cm 7.5cm 1cm},clip]{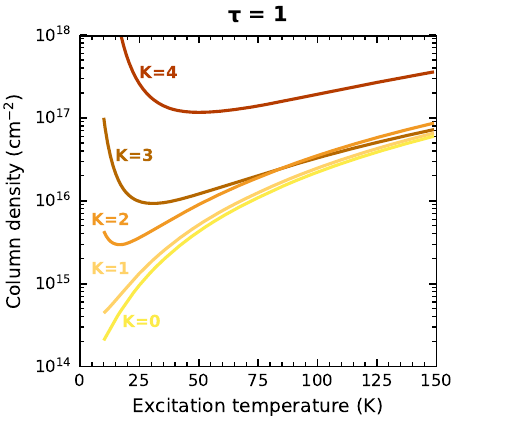}}
\hfill
\subfloat{\includegraphics[trim={0.3cm 14.8cm 7.5cm 1cm},clip]{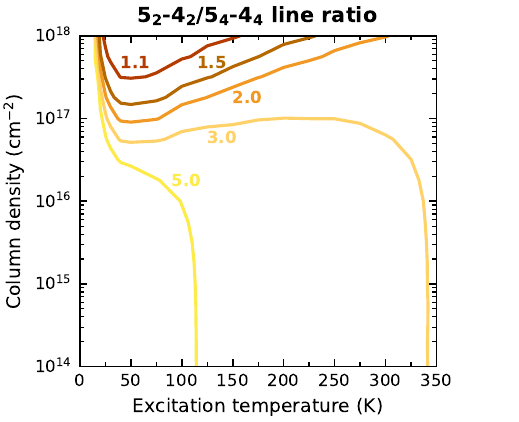}}
\caption{$\tau$=1 contours for the different $K$-components of the CH$_3$CN $J=5-4$ ladder (left panel) and the CH$_3$CN $5_2-4_2/5_4-4_4$ line ratio (right panel) for emission in LTE as function of column density and excitation temperature. A line width of 8.4 km s$^{-1}$ has been adopted (as observed toward Per-emb-17). Observed $5_2-4_2/5_4-4_4$ line ratios range between $\sim$1.5--3.0 for the sources in our sample.}
\label{fig:tau_CH3CN5-4}
\end{figure*}

\begin{figure*}
\centering
\includegraphics[trim={0.2cm 14.3cm 0cm 0.5cm},clip]{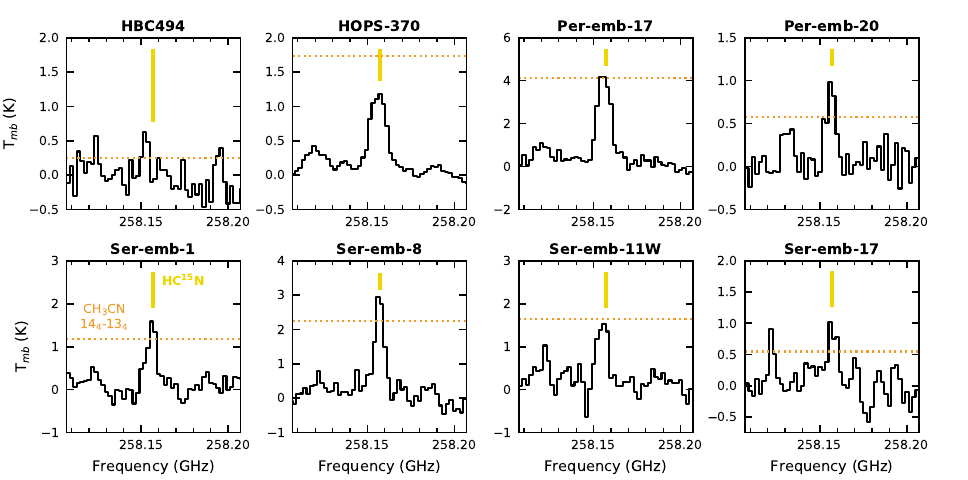}
\caption{Spectra of the HC$^{15}$N $J=3-2$ transition toward the sources in our sample. The line frequency is marked with a yellow vertical line, and the horizontal dotted orange line marks the peak brightness temperature of the CH$_3$CN $14_4-13_4$ transition in that source. }
\label{fig:HC15N}
\end{figure*}

Figure~\ref{fig:tau_CH3CN5-4} presents the $\tau$ = 1 contours for the $J_K=5_K-4_K$ transitions as well as the $5_2-4_2/5_4-4_4$ line ratio as function of column density and excitation temperature for emission in LTE. In addition, Fig.~\ref{fig:HC15N} presents NOEMA spectra for the HC$^{15}$N $J=3-2$ transition toward the sources in our sample as obtained in the 1 mm observations.

\end{appendix}

\end{document}